# A mixed leading variable physics-informed neural network for nonsmooth dynamics simulation in contact systems

Xinyuan Li[a, 1], Zilin Li[a, 1, *], Hongtao Wei[a, *], Ronghan Wei[a, b]

[a] School of Mechanics and Safety Engineering, Zhengzhou University, Zhengzhou, 450001, China

[b] Henan Province Engineering Technology Research Center of MEMS Manufacturing and Applications, Zhengzhou University, Zhengzhou, 450001, China

* Corresponding authors. E-mail addresses: lizilin@zzu.edu.cn (Z. L), htwei@zzu.edu.cn (H. W)

[1] Xinyuan Li and Zilin Li contributed equally to this work.

ABSTRACT

High-accuracy computation of nonsmooth dynamics is a challenging problem in systems with multiple frictional contacts. This study proposes a general physics-informed neural network framework, termed the mixed leading variable physics-informed neural network (MLV-PINN). By using the governing equations and contact-friction laws as physical constraints, this framework provides common solution approaches for computing nonsmooth dynamics not only in discrete systems but also in continuum systems. Specifically, the network represents accelerations, normal contact forces and friction margins as independent outputs and determines them simultaneously, without first reducing the dynamics to discrete velocity-impulse complementary subproblems. A composite physics-informed loss is constructed from the equations of motion in their original form, contact-friction complementary conditions and a friction-margin consistency relation. To improve stick-slip state identification during PINN optimisation, a mixed leading variable combining relative tangential velocity and acceleration is introduced to reformulate the tangential complementary conditions. Four benchmark simulations cover one-dimensional (1-D) nonsmooth dynamics involving single-contact stick-slip, 1-D nonsmooth dynamics involving multi-contact stick-slip, two-dimensional (2-D) nonsmooth dynamics with coupled normal contact-separation and tangential stick-slip, and 2-D nonsmooth dynamics in a continuum system involving moving contact on an elastic beam. Benchmark comparisons show substantial reductions in displacement and velocity errors relative to the corresponding conventional methods at large time steps. These results, together with an ablation study and further tests involving friction-velocity slopes, long-time integration and loss weights, illustrate the computational accuracy and support the broad applicability of the proposed framework.

## 1. Introduction

Contact and friction are key interfacial mechanisms that induce nonsmooth state transitions and complex dynamical behaviour in mechanical systems [1, 2]. Nonsmooth dynamics problems involving contact and friction are widespread in engineering systems, including high-speed train braking systems [3], robotic joints [4], and rotating machinery [5], for which accurate modelling and computation are essential to dynamic performance and operational stability. Numerical solution methods for nonsmooth multibody dynamics have long attracted considerable attention [6]. In nonsmooth multibody dynamics, tangential stick-slip and normal contact-separation transitions constitute two typical classes of interfacial state transitions. Unlike smooth dynamical systems, the interfacial states cannot be determined a priori but must be solved simultaneously with the dynamic response. For complex systems containing a large number of contact points, the rapid growth in possible interfacial-state combinations makes accurate and efficient computation of state-dependent dynamic responses particularly challenging [7, 8].

Unilateral contact problems involving normal contact and friction are commonly formulated as complementary relations or variational inequalities and can, under appropriate conditions, be recast as equivalent constrained optimisation problems [9, 10]. For the resulting nonsmooth dynamical systems, existing time-integration methods can be broadly classified into event-driven [9, 11] and time-stepping [12, 13] methods. Event-driven methods explicitly locate the instants of interfacial state transitions, integrate the piecewise-smooth equations of motion between successive events, and update the state variables according to the corresponding transition laws. By contrast, time-stepping methods do not explicitly track individual events but jointly solve the system states and contact forces or impulses over discrete time intervals. For dynamic simulation in systems with a large number of contact points, event-driven solvers rely on explicit event detection and become impractical for systems with many contacts, whereas time-stepping solvers are intrinsically limited in accuracy at large time steps.

Recent studies have sought to improve the modelling generality, time-integration accuracy, and computational efficiency of nonsmooth solvers through unified formulations of nonsmooth constraints [14], high-order time-integration schemes [15], and computationally efficient algorithms for multi-contact problems [16], respectively. Regarding the unified formulation of nonsmooth constraints, Kan et al. [14] employed a modified Fischer-Burmeister function to reformulate frictional sticking-sliding conditions as a unified system of nonlinear complementary equations, avoiding separate solution formulations for the sticking and sliding states. For high-order time integration, Capobianco et al. [15] proposed Lobatto-type variational integrators that extend high-order temporal discretisation with internal stages to problems involving unilateral contact and set-valued Coulomb friction. Wang et al. [17] formulated frictional contact in flexible multibody systems as a cone complementary problem and combined it with the generalised-α method to solve large-deformation contact dynamics. For multi-contact complementary problems, Fusai and Tasora [16] introduced Anderson acceleration into a fixed-point iteration framework to solve cone complementary problems arising in multibody systems with numerous bilateral constraints and nonsmooth frictional contacts, thereby improving the convergence efficiency of the complementary iterations. These developments point to several possible routes towards more accurate and efficient computation of nonsmooth dynamics.

Recently, physics-informed neural networks (PINNs) have been shown to provide a physics-constrained machine learning framework for solving forward and inverse problems governed by differential equations [18, 19]. A classical PINN typically operates by feeding spatio-temporal coordinates and system parameters into a neural network to predict displacements, velocities, or other physical fields, while automatic differentiation enables the governing equations, initial and boundary conditions, and observational information to be incorporated into the loss function [20, 21]. To address problems involving temporal evolution, researchers have further developed approaches such as discrete-time PINNs [21], space-time domain decomposition [22], and causal training [23]. Jagtap and Karniadakis [22] proposed the extended physics-informed neural network (XPINN), which decomposes the space-time domain into multiple subdomains, constructs an independent subnetwork for each subdomain, and couples these subnetworks through interface conditions. Wang et al. [23] introduced temporal causal weights into the loss function, directing the network to prioritise the reduction of governing-equation residuals over earlier time intervals before progressively learning the subsequent temporal evolution.

In other areas of computational mechanics, researchers have developed both strong-form formulations based directly on the governing equations and variational frameworks, such as the deep Ritz method and the deep energy method [24-26]. The former incorporate the residuals of the local governing equations into the loss function and are commonly referred to as strong-form PINNs. This formulation facilitates the joint representation of stresses, internal forces and other physical quantities. By contrast, energy-based formulations can reduce the order of the derivatives required in the loss function and are therefore better suited to solid mechanics problems with a well-defined variational structure. Hard enforcement of boundary conditions, adaptive sampling and dynamic loss weighting have also been employed to improve the training process [27-29]. Although these methods and strategies were not developed specifically for nonsmooth dynamics, the resulting advances in the formulation of physical constraints and training optimisation provide useful guidance for applying PINNs to contact dynamics problems.

However, extending PINNs from problems involving smooth dynamics to contact problems requires the treatment of unilateral constraints and nonsmooth transitions between interface states. One straightforward strategy is to introduce a penalty term for the unilateral contact constraint into the loss function. However, selecting the penalty coefficient requires a balance between constraint accuracy and the conditioning of the optimisation problem [30]. More recently, studies have begun to explore the direct representation of contact complementary in PINNs using mixed-variable formulations and nonlinear complementary functions. Sahin et al. [31] used displacement and stress as mixed network outputs, imposed selected boundary conditions through output transformations, and converted the normal-contact KKT conditions into complementary residuals that can be minimised during training. The resulting framework was applied to forward and inverse contact problems, as well as parameter identification. In subsequent work, Sahin et al. [32] extended the PINN framework based on displacement-stress mixed variables and complementary residuals to three-dimensional small-deformation contact problems. A separate line of research is based on energy principles. Bai et al. [33] represented large-deformation frictionless contact using a surface contact potential and employed relaxation, incremental loading and output scaling to improve nonlinear training. These studies demonstrate that unilateral contact constraints can be transformed into physical constraints suitable for neural-network optimisation through penalty functions, complementary residuals or contact-energy terms.

For the computation of nonsmooth dynamics in frictional systems, Li et al. [34] embedded multibody time-stepping equations and linear complementary problem (LCP) conditions into network training and developed four PINN frameworks, demonstrating the feasibility of combining complementary constraints with PINNs. However, replacing conventional solvers with PINNs while retaining the discretised complementary formulation does not fundamentally overcome its accuracy limitations [34-36]. Although instantaneous velocity or acceleration can define the interface state in principle, criteria based on a single instantaneous quantity are sensitive to numerical errors and cannot robustly identify state transitions. Additional state variables have therefore been introduced to describe interface evolution. Wang et al. [37] incorporated a rate-and-state friction model into PINNs, whereas Borate et al. [38] used transmitted-wave amplitudes as proxies for friction states to identify stick-slip parameters. However, these variables are used primarily for frictional constitutive modelling or state identification. PINN-based nonsmooth dynamic computation still requires the development of a tangential criterion variable that captures the evolution trend of the motion and can be incorporated into the optimisation process.

For the computation of nonsmooth dynamics in continuum systems, a typical procedure is to discretise the spatial domain using finite element or mortar methods and then apply nonsmooth time integration to solve the resulting contact-friction complementary problems or variational inequalities [17, 39, 40]. In such formulations, the coupling among continuum dynamic fields, time-varying contact regions and high-dimensional complementary constraints after discretisation makes computational accuracy and efficiency sensitive to mesh resolution, contact-search procedures, time-step size and the number of active contact constraints. Contact PINN formulations have been reported for static, frictionless continuum problems [32, 33]. Nonsmooth dynamic PINNs capable of representing stick-slip and contact-separation transitions have also been demonstrated for multibody systems with a finite number of degrees of freedom [34, 41]. However, relatively few PINN frameworks have been reported for multi-point frictional contact in continuum systems. Developing such frameworks offers a promising route for jointly resolving spatio-temporal continuum responses and interface-state transitions.

This study proposes a generic mixed leading variable physics-informed neural network (MLV-PINN) framework for high-accuracy computation of nonsmooth dynamics in frictional systems. A physics-informed loss function is constructed from the original equations of motion and contact-friction complementary laws, avoiding their reduction to discrete velocity-impulse complementary subproblems. The tangential complementary conditions are reformulated using a mixed leading variable that combines relative tangential velocity with information on its evolution within the time step. A network architecture is also designed with accelerations, normal contact forces and friction margins as independent outputs, enabling motion responses, interface forces and contact states to be determined jointly under unified physical constraints. Implicit Runge-Kutta (IRK) integration is incorporated into this architecture to provide high-order state updates. The resulting framework is applicable not only to discrete systems with multiple contact points but also to continuum and moving-contact systems. This paper is organised as follows: Section 2 introduces the contact-friction system and the MLV-PINN framework. Section 3 presents numerical examples and comparative results for one-dimensional stick-slip, two-dimensional coupled stick-slip and contact-separation, and nonsmooth continuum systems. Section 4 examines leading-variable ablation, steep negative friction-velocity slopes, long-time integration, and training configurations.

## 2. MLV-PINN framework

This section presents the overall architecture of the MLV-PINN framework for nonsmooth multibody dynamics. The framework constructs its physics-informed formulation based on the equations of motion and the normal-contact and tangential-friction complementary constraints, which are incorporated into a unified loss function for optimisation. In addition, to address the computational difficulties associated with contact-state transitions in frictional systems, the tangential complementary conditions at the contact interface are reformulated to improve the sensitivity of the numerical scheme to state-transition boundaries.

In the following, the dynamic and unilateral-contact theory of multibody systems are first presented, followed by the construction of the mixed leading variable and the reformulation of the tangential complementary relation. Then, the MLV-PINN framework is developed by combining neural-network training, a smoothed Fischer-Burmeister complementary mapping, a composite physics-informed loss function, and implicit Runge-Kutta time discretisation.

*2.1. Dynamics and unilateral constraint formulations of multibody systems*

Fig. 1 schematically illustrates a contact system involving a normal gap and relative tangential sliding. Fig. 1(a) shows the two bodies in a separated state, with the normal gap at the interface denoted by $g_{\mathrm{N}}$. For example, $g_{\mathrm{N}} > 0$ represents that the normal contact constraint is inactive. In Fig. 1(b), the two bodies are in contact and the normal gap reduces to $g_{\mathrm{N}} = 0$. Fig. 1(c) shows an enlarged view of the local contact region. When the contact interface is represented by multiple contact points, the blue and yellow points denote paired interaction points on the two opposing surfaces. Each contact pair is associated with a normal contact force $\boldsymbol{\lambda}_{\mathrm{N}}$ and a tangential friction force $\boldsymbol{\lambda}_{\mathrm{T}}$. The tangential friction force $\boldsymbol{\lambda}_{\mathrm{T}}$ always opposes the local tangential relative motion $\gamma_{\mathrm{T}}$ in the sliding state, or the tendency towards relative motion in the sticking state. Fig. 1(d) presents a velocity-dependent friction law (top) and a representative phase trajectory of friction-induced stick-slip oscillation (bottom), illustrating the switching behaviour between sticking and sliding. Here, $\gamma_{\mathrm{T}}$, $q$, and $v$ denote the tangential relative velocity, displacement, and velocity, respectively. As shown in Fig. 1, normal contact-separation transitions and friction-induced tangential stick-slip transitions constitute the two principal sources of nonsmoothness in the system.

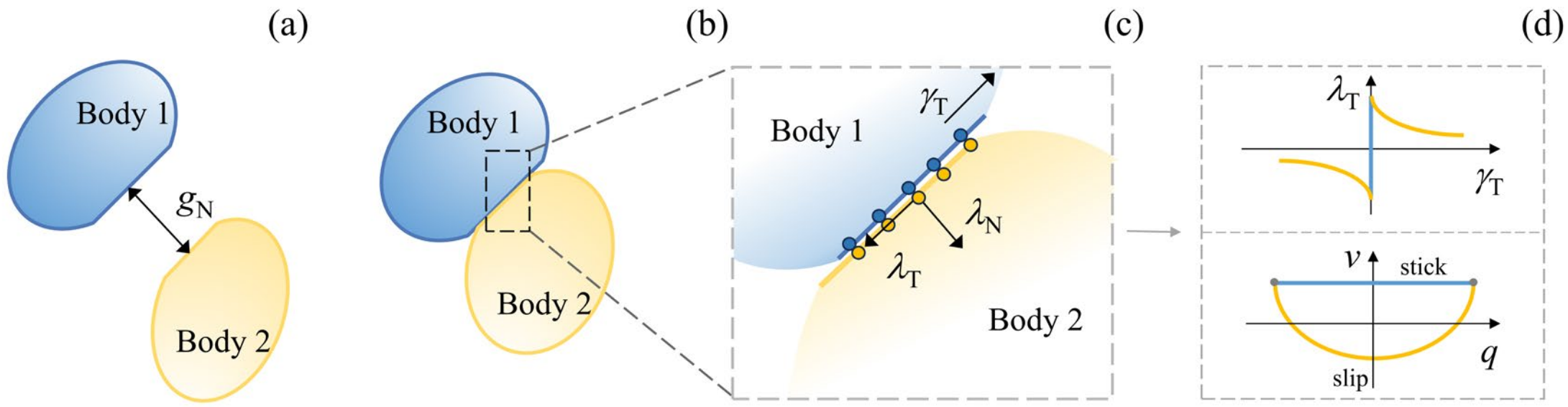

Fig. 1. Schematic of a contact-friction system: (a) separated state; (b) contact state; (c) frictional contact region; (d) velocity-dependent friction law and nonsmooth phase trajectory associated with stick-slip transition.

For the dynamic modelling of multibody systems with clearance, the normal contact force $\boldsymbol{\lambda}_{\mathrm{N}}$ and tangential friction force $\boldsymbol{\lambda}_{\mathrm{T}}$ acting at the contact interfaces must be taken into account. Let $n$ denote the number of degrees of freedom of the system. The generalised displacement, velocity, and acceleration are denoted by $\mathbf{q}(t) \in \mathbb{R}^n$ , $\mathbf{v}(t) = \dot{\mathbf{q}}(t) \in \mathbb{R}^n$ , $\mathbf{a}(t) = \ddot{\mathbf{q}}(t) \in \mathbb{R}^n$ , respectively. The equations of motion can then be written as:

$$\mathbf{M}(\mathbf{q})\ddot{\mathbf{q}} - \mathbf{h}(\mathbf{q}, \dot{\mathbf{q}}, t) - \mathbf{W}_{\mathrm{N}}\boldsymbol{\lambda}_{\mathrm{N}} - \mathbf{W}_{\mathrm{T}}\boldsymbol{\lambda}_{\mathrm{T}} = 0 \tag{1}$$

where $\mathbf{M}(\mathbf{q})$ denotes the mass matrix and $\mathbf{h}(\mathbf{q}, \dot{\mathbf{q}}, t)$ represents the system generalised-force vector excluding the unilateral contact forces. When multiple contact pairs are present, $\boldsymbol{\lambda}_{\mathrm{N}}$ and $\boldsymbol{\lambda}_{\mathrm{T}}$ denote the vectors collecting the local normal contact forces and tangential friction forces, respectively. $\mathbf{W}_{\mathrm{N}}$ and $\mathbf{W}_{\mathrm{T}}$ are the corresponding generalised force direction matrices.

Unlike in smooth dynamical systems, $\boldsymbol{\lambda}_{\mathrm{N}}$ and $\boldsymbol{\lambda}_{\mathrm{T}}$ in Eq. (1) are not prescribed external loads, but unknown interfacial reaction forces determined jointly by the interfacial gaps, local relative motion, and interfacial states. Therefore, the equations of motion in Eq. (1) alone are insufficient to uniquely determine the dynamic response of the system, and additional complementary constraints for normal contact and tangential friction are required.

For multibody systems with unilateral rigid contact, the normal-contact and tangential-friction states can both be characterised by inequality constraints [30]. Rigid normal contact satisfies the Signorini conditions, according to which the normal gap $\mathbf{g}_{\mathrm{N}}$ and normal contact force $\boldsymbol{\lambda}_{\mathrm{N}}$ satisfy:

$$0 \le \mathbf{g}_{\mathrm{N}} \perp \boldsymbol{\lambda}_{\mathrm{N}} \ge 0 \tag{2}$$

Eq. (2) indicates that, when the interface is separated, $\mathbf{g}_{\mathrm{N}} > 0$ and $\boldsymbol{\lambda}_{\mathrm{N}} = 0$ . When the interface is closed, $\mathbf{g}_{\mathrm{N}} = 0$ , and the normal contact force may be positive.

For an elastic contact model represented by a linear contact spring, finite compression of the contact spring could occur which gives $\mathbf{g}_{\mathrm{N}} < 0$ . Eq. (2) is not suitable to describe the contact law in the elastic contact case. A generalised normal complementary variable is introduced as $\boldsymbol{\Omega}_{\mathrm{N}} = \boldsymbol{\lambda}_{\mathrm{N}} + \mathbf{K}_{\mathrm{c}}\mathbf{g}_{\mathrm{N}}$ [42], where $\mathbf{K}_{\mathrm{c}}$ denotes the contact-stiffness matrix. The normal contact complementary condition can then be written as:

$$0 \le \boldsymbol{\lambda}_{\mathrm{N}} \perp \boldsymbol{\Omega}_{\mathrm{N}} \ge 0 \tag{3}$$

This relation indicates that, when the interface is separated, $\boldsymbol{\lambda}_{\mathrm{N}} = 0$ and $\boldsymbol{\Omega}_{\mathrm{N}} = \mathbf{K}_{\mathrm{c}}\mathbf{g}_{\mathrm{N}} \ge 0$ ; when the interface is in elastic contact, $\boldsymbol{\Omega}_{\mathrm{N}} = 0$ and $\boldsymbol{\lambda}_{\mathrm{N}} = -\mathbf{K}_{\mathrm{c}}\mathbf{g}_{\mathrm{N}} \ge 0$ .

For tangential frictional contact, the friction force-velocity relation can be reformulated into two pairs of complementary variables [8]:

$$0 \le \boldsymbol{\lambda}_{\mathrm{R}} \perp \boldsymbol{\gamma}_{\mathrm{R}} \ge 0,\ 0 \le \boldsymbol{\lambda}_{\mathrm{L}} \perp \boldsymbol{\gamma}_{\mathrm{L}} \ge 0 \tag{4}$$

where $\boldsymbol{\gamma}_{\mathrm{R}}$ and $\boldsymbol{\gamma}_{\mathrm{L}}$ denote the non-negative components of the tangential relative velocity $\boldsymbol{\gamma}_{\mathrm{T}}$ in the two different directions, and can be obtained by Eq. (5):

$$\boldsymbol{\gamma}_{\mathrm{R}} = \frac{1}{2}(|\boldsymbol{\gamma}_{\mathrm{T}}| + \boldsymbol{\gamma}_{\mathrm{T}}), \quad \boldsymbol{\gamma}_{\mathrm{L}} = \frac{1}{2}(|\boldsymbol{\gamma}_{\mathrm{T}}| - \boldsymbol{\gamma}_{\mathrm{T}}) \tag{5}$$

The corresponding friction margins $\boldsymbol{\lambda}_{\mathrm{R}}$ and $\boldsymbol{\lambda}_{\mathrm{L}}$ can be obtained by Eq. (6):

$$\boldsymbol{\lambda}_{\mathrm{R}} = \mu\boldsymbol{\lambda}_{\mathrm{N}} + \boldsymbol{\lambda}_{\mathrm{T}}, \ \boldsymbol{\lambda}_{\mathrm{L}} = \mu\boldsymbol{\lambda}_{\mathrm{N}} - \boldsymbol{\lambda}_{\mathrm{T}} \tag{6}$$

where $\mu$ denotes the coefficient of friction.

For the tangential direction, the velocity-level complementary relation uses the relative tangential velocity $\boldsymbol{\gamma}_{\mathrm{T}}$ as the state-discrimination variable. Theoretically, $\boldsymbol{\gamma}_{\mathrm{T}}$ clearly indicates the sliding direction and is therefore the most commonly used state variable in conventional frictional LCP formulations. However, in dynamic computations, the magnitude of $\boldsymbol{\gamma}_{\mathrm{T}}$ approaches zero near a stick-slip transition. Thus, the instantaneous velocity alone provides the optimisation algorithm with weak information to find out whether the relative motion will continue sliding, approach sticking, or reverse direction.

In addition to the velocity-level complementary relation, unilateral contact with Coulomb friction can also be formulated at the acceleration-force level [8], which provides additional information for determining the interfacial state after a state transition. However, the tangential acceleration-force complementary relation cannot independently determine the interfacial state. It is applicable only when the contact is closed and the tangential relative velocity is zero, and must therefore be used together with the velocity-level condition.

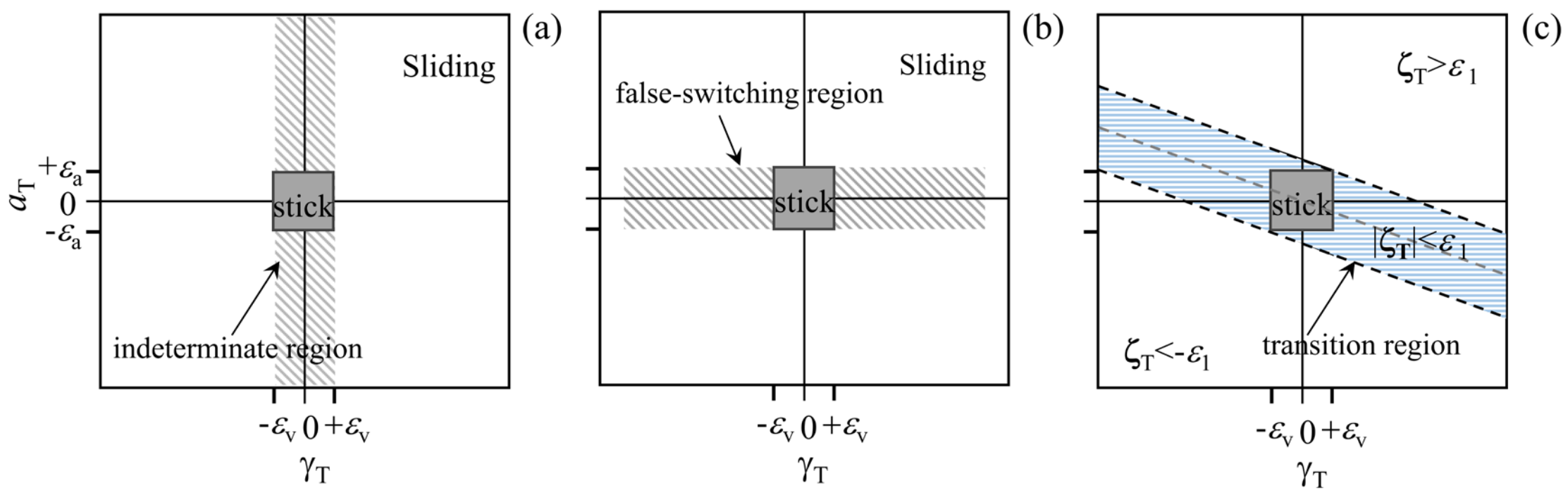

Fig. 2. Schematic comparison of tangential contact-state discrimination criteria in the $(\gamma_{\mathrm{T}}, a_{\mathrm{T}})$ plane: (a) velocity-only criterion; (b) acceleration-only criterion; (c) mixed leading variable (MLV) criterion.

Figs. 2(a)-(b) illustrate the geometric interpretations of the velocity and acceleration criteria in the $(\gamma_{\mathrm{T}}, a_{\mathrm{T}})$ plane, where the horizontal and vertical axes denote the tangential relative velocity $\gamma_{\mathrm{T}}$ and tangential acceleration $a_{\mathrm{T}}$, respectively. As shown in Fig. 2(a), the velocity criterion distinguishes positive and negative sliding according to the sign of $\gamma_{\mathrm{T}}$. The neighbourhood of zero

relative velocity therefore forms a vertical discrimination band, indicating that the criterion depends only on the current relative velocity and does not account for its local evolution. By contrast, the acceleration criterion in Fig. 2(b) partitions the branches according to the sign of $a_{\mathrm{T}}$, resulting in a horizontal discrimination band. Although $a_{\mathrm{T}}$ reflects the direction of velocity variation, its sign does not directly correspond to the actual friction branch. For example, during sliding without a reversal of direction, $a_{\mathrm{T}}$ may change sign while the sign of $\gamma_{\mathrm{T}}$ remains unchanged. Therefore, the state-discrimination variable for the tangential complementary relation should satisfy two requirements: (1) correctly represent the current sliding branch in the sliding regime; and (2) capture the tendency of the relative velocity to approach the zero-velocity boundary in the vicinity of a stick-slip transition.

*2.2. Reformulation of the tangential complementary relation*

This section refines the state-discrimination variable used in the tangential contact complementary relation to improve PINN optimisation near stick-slip transitions. Specifically, rather than using either the tangential relative velocity or acceleration alone as the leading variable, this study constructs a mixed state-discrimination variable tailored to the adopted time-discretisation scheme by combining these two quantities, which is expressed as:

$$\zeta_{\mathrm{T}} = \frac{\gamma_{\mathrm{T}}}{\Delta t} + a_{\mathrm{T}} \tag{7}$$

where $\Delta t$ denotes the discrete time step and $a_{\mathrm{T}} = \dot{\gamma}_{\mathrm{T}}$ is the tangential relative acceleration. Hereafter, $\zeta_{\mathrm{T}}$ is termed the mixed leading variable (MLV). Fig. 2(c) illustrates its operating principle in the $(\gamma_{\mathrm{T}}, a_{\mathrm{T}})$ plane. The regions $\zeta_{\mathrm{T}} > \varepsilon_1$ and $\zeta_{\mathrm{T}} < -\varepsilon_1$ correspond to the two sliding branches, whereas $|\zeta_{\mathrm{T}}| < \varepsilon_1$ defines a transition-sensitive region. Its centreline, $\zeta_{\mathrm{T}} = 0$, satisfies $\gamma_{\mathrm{T}} + \Delta t a_{\mathrm{T}} = 0$, indicating that the current relative velocity can be offset by its local change over the current time step. The resulting transition region is therefore inclined because its location is determined jointly by the current tangential relative velocity and its evolution.

For a contact undergoing unambiguous sliding, the change in relative velocity over one time step is insufficient to bring the velocity to zero, i.e. $|\gamma_T| \gg \Delta t |a_{\mathrm{T}}|$. The velocity term then dominates the MLV, giving $\operatorname{sgn}(\zeta_{\mathrm{T}}) = \operatorname{sgn}(\gamma_{\mathrm{T}})$. Therefore, the MLV retains the positive- or negative-sliding branch identified by the conventional velocity-level LCP relation. As the contact approaches a stick-slip transition, $|\gamma_{\mathrm{T}}|$ becomes comparable to $\Delta t |a_{\mathrm{T}}|$. When $\gamma_{\mathrm{T}} a_{\mathrm{T}} < 0$, the acceleration opposes the current relative tangential motion and moves the state towards the inclined transition region, indicating that the relative velocity may reach or cross zero within the current time step. Within this region, the tangential contact state is determined jointly with the friction-margin conditions: a tangential force within the Coulomb bounds corresponds to sticking, whereas activation of either bound corresponds to sliding in the associated direction.

Subsequently, the tangential complementary relations are reformulated using $\zeta_{\mathrm{T}}$. Specifically, the ReLU function is employed to decompose $\zeta_{\mathrm{T}}$ into two non-negative components, $\zeta_{\mathrm{R}} = \max(0, \zeta_{\mathrm{T}})$ and $\zeta_{\mathrm{L}} = \max(0, -\zeta_{\mathrm{T}})$. These components are then paired with the friction margins $\lambda_{\mathrm{R}}$ and $\lambda_{\mathrm{L}}$ to define the reformulated tangential complementary relations:

$$0 \le \lambda_{\mathrm{R}} \perp \zeta_{\mathrm{R}} \ge 0,\ 0 \le \lambda_{\mathrm{L}} \perp \zeta_{\mathrm{L}} \ge 0 \tag{8}$$

Here, $\lambda_{\mathrm{R}}$ and $\lambda_{\mathrm{L}}$ remain defined by Eq. (6). Therefore, the friction boundary $\lambda_{\mathrm{T}} \in [-\mu\lambda_{\mathrm{N}}, \mu\lambda_{\mathrm{N}}]$ is unchanged. Eq. (8) gives the MLV-based tangential complementary formulation adopted in this work. Section 4.1 further compares the velocity-level, acceleration-level, and MLV-based tangential complementary formulations through an ablation study of stick-slip transition prediction.

*2.3. Physics-informed loss function*

For PINN training, a composite physics-informed loss function is designed as:

$$\mathcal{L}_{\mathrm{total}} = w_{\mathrm{dyn}}\mathcal{L}_{\mathrm{dyn}} + w_{\mathrm{N}}\mathcal{L}_{\mathrm{N}} + w_{\mathrm{T}}\mathcal{L}_{\mathrm{T}} + w_{\mathrm{cons}}\mathcal{L}_{\mathrm{cons}} \tag{9}$$

where $\mathcal{L}_{\mathrm{dyn}}$, $\mathcal{L}_{\mathrm{N}}$, $\mathcal{L}_{\mathrm{T}}$, and $\mathcal{L}_{\mathrm{cons}}$ denote the equations-of-motion residual loss, normal-contact complementary loss, tangential-friction complementary loss, and friction-margin consistency loss, respectively, while $w_{\mathrm{dyn}}$, $w_{\mathrm{N}}$, $w_{\mathrm{T}}$, and $w_{\mathrm{cons}}$ are the corresponding weighting coefficients.

First, the equations-of-motion loss $\mathcal{L}_{\mathrm{dyn}}$ is defined as

$$\mathcal{L}_{\mathrm{dyn}} = \left\| \mathbf{M}\ddot{\mathbf{q}} - \mathbf{h} - \mathbf{W}_{\mathrm{N}}\boldsymbol{\lambda}_{\mathrm{N}} - \mathbf{W}_{\mathrm{T}}\boldsymbol{\lambda}_{\mathrm{T}} \right\|^2 \tag{10}$$

This loss constrains the network-predicted generalised acceleration $\ddot{\mathbf{q}}$, together with the corresponding motion state ($\dot{\mathbf{q}}$, $\mathbf{q}$), normal contact forces $\boldsymbol{\lambda}_{\mathrm{N}}$, and tangential friction forces $\boldsymbol{\lambda}_{\mathrm{T}}$ to satisfy the original second-order equations of motion of the multibody system in Eq. (1).

Next, physics-informed loss terms ($\mathcal{L}_{\mathrm{N}}$ and $\mathcal{L}_{\mathrm{T}}$) are constructed for the normal-contact and tangential-friction complementary constraints. For a pair of complementary variables $\chi$ and $\psi$, the complementarity conditions require $\chi \ge 0$, $\psi \ge 0$, and $\chi\psi = 0$. A straightforward approach is to enforce these requirements separately by penalising negative values of $\chi$ and $\psi$ and a nonzero product $\chi\psi$. However, this introduces multiple penalty terms and associated loss weights. Moreover, near a state-transition boundary, both complementary variables may approach zero, causing the product residual $\chi\psi$ and its gradients to become small and thus provide only weak constraint information for optimisation. To avoid this difficulty, the Fischer-Burmeister (FB) function is employed to transform the complementary conditions into equality residuals [31]. The smoothed FB formulation adopted in the optimisation is given by:

$$\phi_\epsilon(\chi,\psi) = \chi + \psi - \sqrt{\chi^2 + \psi^2 + \varepsilon} \tag{11}$$

where $\varepsilon > 0$ is the smoothing parameter. For finite $\varepsilon$, Eq. (11) provides a smooth approximation to the standard FB mapping, and the complementary constraints are imposed approximately by minimising the corresponding residuals. To reduce the influence of differences in physical dimensions and numerical scales on the optimisation, each physical residual is nondimensionalised using its corresponding characteristic scale. Based on Eq. (11), the normal-contact complementary loss $\mathcal{L}_{\mathrm{N}}$ and tangential-friction complementary loss $\mathcal{L}_{\mathrm{T}}$ are defined as:

$$\mathcal{L}_{\mathrm{N}} = \left\| \phi_\epsilon(\boldsymbol{\lambda}_{\mathrm{N}}, \boldsymbol{\Omega}_{\mathrm{N}}) \right\|^2 \tag{12}$$

$$\mathcal{L}_{\mathrm{T}} = \sum_{\alpha\in\{\mathrm{R},\mathrm{L}\}} \left\| \phi_\epsilon\left(\zeta_\alpha, \boldsymbol{\lambda}_\alpha\right) \right\|^2 \tag{13}$$

where $\mathcal{L}_{\mathrm{N}}$ corresponds to the normal contact-separation complementary relation, and $\mathcal{L}_{\mathrm{T}}$ corresponds to the tangential-friction complementary relation reformulated using the MLV. Fig. 3 illustrates the relationship among the ReLU-based branch decomposition, the standard FB function, and the smoothed FB function.

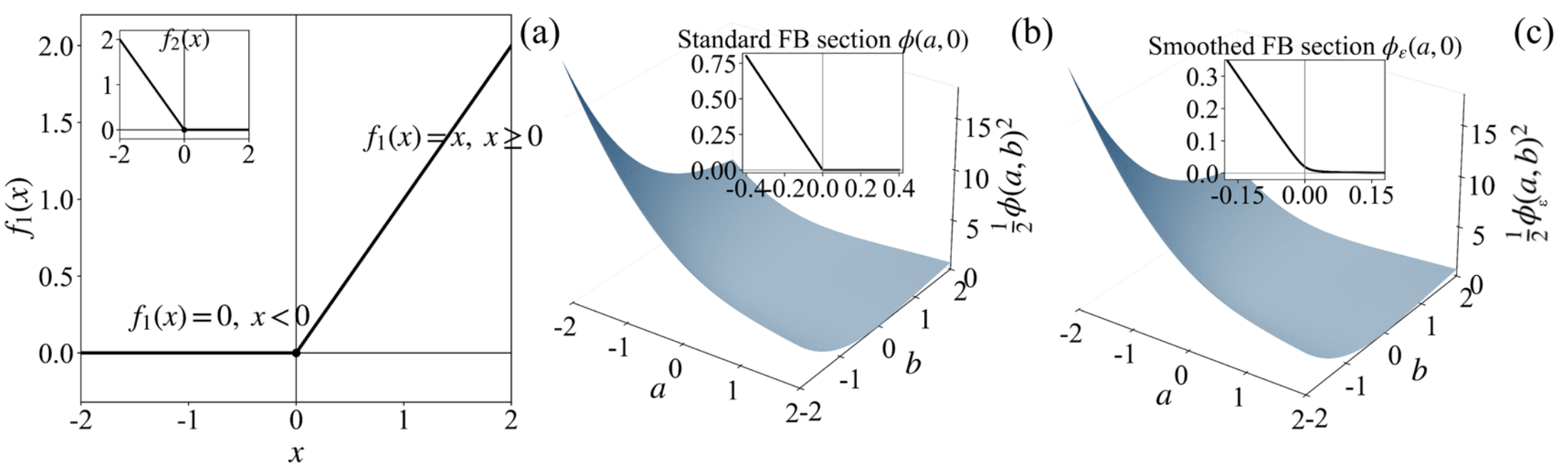

Fig. 3. Transformation of nonsmooth complementary constraints into differentiable residuals: (a) ReLU-based branch decomposition; (b) standard Fischer-Burmeister function; (c) smoothed Fischer-Burmeister function.

Finally, since the normal contact force $\boldsymbol{\lambda}_{\mathrm{N}}$ and the two friction margins $\boldsymbol{\lambda}_{\mathrm{R}}$ and $\boldsymbol{\lambda}_{\mathrm{L}}$ are treated as independent variables in the formulation, the relations defined in Eq. (6) are not automatically satisfied. A friction-margin consistency loss $\mathcal{L}_{\mathrm{cons}}$ is therefore introduced to enforce these relations:

$$\mathcal{L}_{\text{cons}} = \left\| \boldsymbol{\lambda}_{\text{R}} + \boldsymbol{\lambda}_{\text{L}} - 2\mu\boldsymbol{\lambda}_{\text{N}} \right\|^2 \tag{14}$$

*2.4. Architecture of MLV-PINN*

Fig. 4 presents the overall architecture of the proposed MLV-PINN. In Step 1, the system parameters and the motion state at the beginning of the current time step are supplied to the network to predict the response over the interval $[t_n, t_n + \Delta t]$. In Step 2, the network simultaneously predicts the accelerations, normal contact forces, and friction margins associated with the positive and negative directions at all IRK stages within the current time step. Notably, the normal forces and friction margins are treated as independent network outputs rather than being calculated through discretised dynamic relations.

To reconstruct velocities and displacements from the predicted accelerations while maintaining their kinematic relations, a high-order Gauss-Legendre IRK scheme is employed. Using the known state at the beginning of the time step, the velocities and displacements at the collocation points are reconstructed from the corresponding stage accelerations. These operations are shown in Steps 3(a) and 3(b) of Fig. 4, and the corresponding integration relations are given in Eq. (15):

$$\mathbf{v}_n^{(i)} = \mathbf{v}_n + \Delta t \sum_{j=1}^{s} A_{ij} \mathbf{a}_n^{(j)}, \ \mathbf{q}_n^{(i)} = \mathbf{q}_n + \Delta t \sum_{j=1}^{s} A_{ij} \mathbf{v}_n^{(j)} \tag{15}$$

where $n$ denotes the time-step index, $s$ is the number of IRK stages, and $i, j = 1, \ldots, s$ are the IRK stage indices, where $i$ denotes the current stage and $j$ is the summation index. The $i$th IRK stage corresponds to the $i$th Gauss-Legendre collocation point $t_n + c_i \Delta t$ within the current time interval $[t_n, t_{n+1}]$, where $c_i$ specifies the location of the collocation point within the time step. $A_{ij}$ is the $(i, j)$ entry of the IRK coefficient matrix, and $\mathbf{a}_n^{(j)}$ denotes the acceleration at the $j$th IRK stage of time step $n$, i.e. at the corresponding Gauss-Legendre collocation point.

In addition to reconstructing the motion state within the time interval, Steps 3(c)-(f) use the network outputs and the reconstructed kinematic variables to construct the auxiliary variables required for the loss terms in Eq. (9). Step 3(c) determines the tangential friction force from the two friction margins $\lambda_{\text{R}}$ and $\lambda_{\text{L}}$. Step 3(d) constructs the normal complementary variable $\Omega_{\text{N}}$ from the normal contact force and the normal gap. Step 3(e) constructs the mixed leading variable $\zeta_{\text{T}}$ from the tangential relative velocity and acceleration at each IRK stage. Step 3(f) then decomposes this variable into its positive- and negative-direction non-negative components to formulate the corresponding tangential complementary residuals. The MLV is therefore evaluated from the local tangential motion state at each IRK stage and used to identify the friction branch at that stage.

In Step 4, the loss terms defined in Section 2.3 are evaluated at each IRK stage and averaged over the stages to form the composite physics-informed loss for the current time step. After the network parameters have been optimised in Step 5, Step 6 updates the displacement and velocity at the end of the current time step using the IRK formulations. The updated state is then used as the initial condition for the next time step. The complete solution procedure is summarised in Algorithm 1.

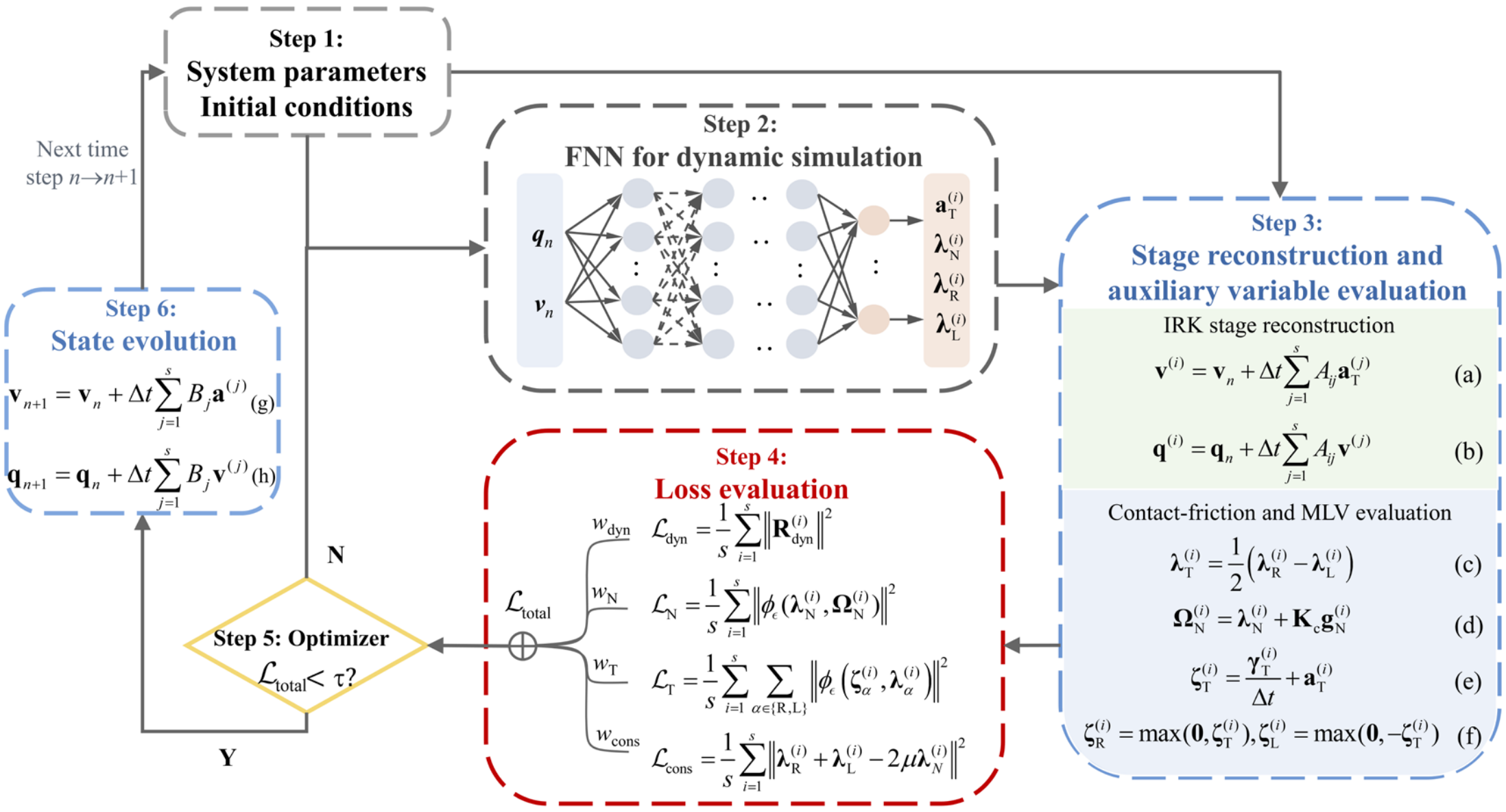

Fig. 4. Architecture and computational workflow of the MLV-PINN. For clarity, the time-step subscript $n$ is omitted from the IRK-stage variables; superscript $i$ or $j$ denotes stages within the current time interval $[t_n, t_{n+1}]$.

**Algorithm 1**. Stepwise solution procedure of MLV-PINN

*Input*: System parameters, initial state vector $\{\mathbf{q}_0, \mathbf{v}_0\}$, time-step size $\Delta t$, total number of time steps $N_t$, and Gauss-Legendre IRK coefficients $A_{ij}$, $B_i$, and $c_i$ $(i, j = 1, \ldots, s)$.

*Output*: Time histories of the state vector $\{\mathbf{q}_{n+1}, \mathbf{v}_{n+1}\}_{n=0}^{N_t-1}$, together with the corresponding normal contact-force and tangential friction-force histories.

*Step 1*: Initially, set $n$=0. At the current time step $n$, feed the system parameters and the system state $\{\mathbf{q}_n, \mathbf{v}_n\}$ at $t_n$ into the network.

*Step 2*: Predict the unknown quantities at all IRK stages, including the stage accelerations $\mathbf{a}_n^{(i)}$, normal contact forces $\boldsymbol{\lambda}_{N,n}^{(i)}$, and positive- and negative-direction tangential friction margins $\boldsymbol{\lambda}_{R,n}^{(i)}$ and $\boldsymbol{\lambda}_{L,n}^{(i)}$.

*Step 3*: Calculate the stage velocities $\mathbf{v}_n^{(i)}$ and displacements $\mathbf{q}_n^{(i)}$ according to the IRK algebraic relations shown in Steps 3(a)-(b) of Fig. 4. The corresponding tangential friction forces $\boldsymbol{\lambda}_{T,n}^{(i)}$, normal complementary variables $\boldsymbol{\Omega}_{N,n}^{(i)}$, mixed leading variables $\zeta_{T,n}^{(i)}$, and their positive and negative components $\zeta_{R,n}^{(i)}$ and $\zeta_{L,n}^{(i)}$ are then calculated according to Steps 3(c)-(f) of Fig. 4.

*Step 4*: Evaluate the physical loss terms defined in Section 2.3 at all IRK stages and average them over the stages to form the composite physics-informed loss $\mathcal{L}_{total}$ for the current time step.

*Step 5*: Update the network parameters using the optimiser and repeat Steps 2-4 until the prescribed convergence criterion for the total loss is satisfied.

*Step 6*: Update the system state according to the end-of-step IRK integration relations shown in Steps 6(g)-(h) of Fig. 4 to obtain $\{\mathbf{q}_{n+1}, \mathbf{v}_{n+1}\}$. Set $n \leftarrow n+1$ and repeat Steps 1-6 until all time steps are completed.

*Remark*: $\mathbf{R}_{\text{dyn}}^{(i)}$ in Fig. 4 denotes the equations-of-motion residual evaluated at the *i*th IRK stage, whose squared norm defines $\mathcal{L}_{\text{dyn}}$ in Eq. (10).

MLV-PINN treats the accelerations, normal contact forces, and friction margins as independent network outputs and uses a high-order numerical integration scheme to reconstruct the displacements and velocities from the predicted stage accelerations. The motion and contact variables are thus determined jointly under the governing equations and contact-friction constraints. This network design reduces the reliance of contact-force determination on discretised dynamic relations and is intended to limit error propagation among physical quantities during contact-state transitions, thereby mitigating error accumulation in long-time simulations.

## 3. Numerical validation and discussion

This section validates MLV-PINN through four representative nonsmooth dynamics problems of increasing complexity. Case I uses the classical single-slider-belt model to assess the ability of MLV-PINN to capture tangential stick-slip transitions. Case II introduces coupled vibration through a lumped-parameter multi-slider-belt model, while retaining tangential friction as the source of nonsmoothness. It assesses the prediction of stick-slip transitions and dynamic responses in a coupled multi-degree-of-freedom system. Case III uses a classical friction-induced self-excited vibration model with coupled transverse and longitudinal motion. By additionally considering normal contact-separation transitions, this case examines the capability of MLV-PINN to resolve coupled normal-tangential nonsmooth dynamics. Case IV extends the method to a continuum frictional system to assess its applicability to nonsmooth vibration analysis with moving contact locations and spatially coupled contact-friction behaviour. Together, these cases provide a systematic assessment of the computational accuracy, state-transition identification capability, and applicability of MLV-PINN across increasingly complex systems.

The neural network is a fully connected feedforward network with five hidden layers and 40 neurons per hidden layer, using the Tanh activation function. Training proceeds in two stages, using Adam followed by L-BFGS-B. The learning rate decays from $2\times10^{-3}$ to $10^{-5}$, the smoothing parameter is set to $\varepsilon=10^{-9}$, and the maximum number of L-BFGS-B iterations is 2000. The network parameters are reinitialised at each time step. Each residual term is nondimensionalised using its corresponding characteristic scale. The effects of the loss weights and the number of Adam pre-training steps are discussed in Section 4.4. Unless otherwise stated, all cases use the same network architecture and optimisation strategy.

*3.1. Friction models*

The law of friction is an important factor for nonsmooth behaviour. Two Stribeck friction models are adopted [43, 44] to simulate stick-slip oscillations. Both describe a decrease in the friction coefficient with increasing relative sliding speed, but differ in their decay profiles and limiting values. The first model is expressed in rational form:

$$\mu = \frac{\mu_0}{1+\delta\left|\gamma_{\mathrm{T}}\right|} \tag{16}$$

and the second takes an exponential form:

$$\mu = \mu_k + (\mu_s - \mu_k)e^{-\alpha\left|\gamma_{\mathrm{T}}\right|} \tag{17}$$

where $\mu_0$ denotes the reference friction coefficient at zero relative sliding velocity, $\mu_s$ and $\mu_k$ are the static and kinetic friction coefficients, respectively, and $\delta$ and $\alpha$ are the friction-velocity decay parameters.

*3.2. Case I: One-dimensional nonsmooth simulation of a classical stick-slip benchmark*

*3.2.1. Physical problem and formulation*

Fig. 5 shows a classical slider-belt friction system, which is commonly used as a benchmark for evaluating numerical methods for nonsmooth dynamics. The nonsmoothness of this system arises solely from stick-slip transitions at the contact interface, without normal separation, impact, or reattachment. It therefore provides a benchmark for assessing the ability of MLV-PINN to distinguish between sticking and sliding states.

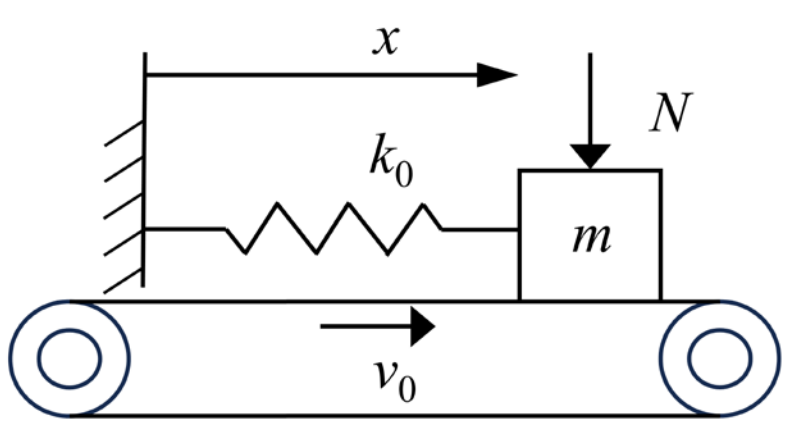

Fig. 5. Mechanical model of Case I: a classical slider-belt system exhibiting stick-slip motion.

In the model shown in Fig. 5, a slider of mass $m$ is connected to the fixed support on the left by a linear spring of stiffness $k_0$ and is in contact with a belt moving at a constant velocity $v_0$ under a normal load $N$. The system has a single degree of freedom, represented by the horizontal displacement $x$ along the direction of belt motion. The friction force between the slider and the belt is assumed to follow a Stribeck friction law. When the slider remains stationary relative to the belt, static friction maintains sticking, provided that the force required to maintain sticking remains within the static friction limit. Once this required force exceeds the limit, relative sliding occurs, and the friction force depends on the relative sliding speed. The resulting switching between sticking and sliding states constitutes the tangential nonsmooth behaviour considered in this case.

Let $x(t)$ denote the absolute displacement of the slider. The equation of motion is written as

$$m\ddot{x} + k_0 x - \lambda_{\mathrm{T}} = 0 \tag{18}$$

where the friction force $\lambda_{\mathrm{T}}$ is determined jointly by the velocity-dependent friction model in Eq. (16) and the tangential complementary relation. In MLV-PINN, the residual of Eq. (18) is evaluated at each IRK stage to construct the equations-of-motion residual loss $\mathcal{L}_{\mathrm{dyn}}$. This loss is combined with the tangential complementary loss $\mathcal{L}_{\mathrm{T}}$ in Eq. (13) and the friction-margin consistency loss $\mathcal{L}_{\mathrm{cons}}$ in Eq. (14) to form the physics-informed loss for the current time step.

*3.2.2. Results and discussion*

In this case, the network inputs are the displacement and velocity at the beginning of the current time step, while the network outputs are the acceleration $\mathbf{a}^{(i)}$ and the positive- and negative-direction friction margins $\boldsymbol{\lambda}_{\mathrm{R}}^{(i)}$ and $\boldsymbol{\lambda}_{\mathrm{L}}^{(i)}$ at each IRK stage. To validate the computational results, MLV-PINN is compared with the benchmark result and the conventional LCP method. The benchmark result is obtained using the Switch Model with a MATLAB ODE solver and a small time step of $\Delta t = 10^{-4}$ to ensure high numerical accuracy. The conventional LCP method employs an LCP-based time-stepping scheme with a time step of $\Delta t = 10^{-2}$. MLV-PINN uses the same time-step size, $\Delta t = 10^{-2}$. The model parameters are $m = 1.0$, $k_0 = 100$, $c_0 = 0$, $v_0 = 1.0$, and $N = 1000$. The friction model is given by Eq. (16), with $\mu_0 = 0.1$ and $\delta = 2$. The initial conditions are $x(0) = 1.0$ and $\dot{x}(0) = 1.0$.

Figs. 6(a)-6(d) compare the displacement, velocity, and friction-force time histories and the displacement-velocity phase trajectories obtained by the three methods. The black solid line with filled-circle markers denotes the benchmark result, the blue dashed line with rectangular markers denotes the conventional LCP method, and the red short-dashed line with upward-pointing triangular markers denotes MLV-PINN. As shown in Fig. 6(a), the displacement predicted by MLV-PINN agrees with the benchmark result over multiple oscillation cycles. By contrast, the conventional LCP method exhibits a phase shift after approximately 5 s, after which its displacement response gradually deviates from the benchmark result.

The velocity response more directly reveals how accurately each method captures stick-slip transitions. As shown in Fig. 6(b), the velocity predicted by MLV-PINN returns to the sticking velocity at the end of sliding and agrees with the benchmark result during sticking. The conventional

LCP method shows a clear transition delay from the third oscillation cycle, with the velocity still deviating from the sticking condition when the benchmark result enters the sticking stage. The friction-force response in Fig. 6(c) also shows this transition delay. Apparently, the MLV-PINN prediction is closer to the benchmark result.

Fig. 6(d) shows the displacement-velocity phase trajectories. In the benchmark result, the sticking stages appear as approximately horizontal segments at the sticking velocity. MLV-PINN preserves this structure and its connection to the sliding trajectories, whereas the conventional LCP method exhibits trajectory deviations in the stick-slip transition region. Together with the time histories in Figs. 6(a)-6(c), these results show that MLV-PINN captures tangential stick-slip transitions and the resulting dynamic response more accurately at the time step used here.

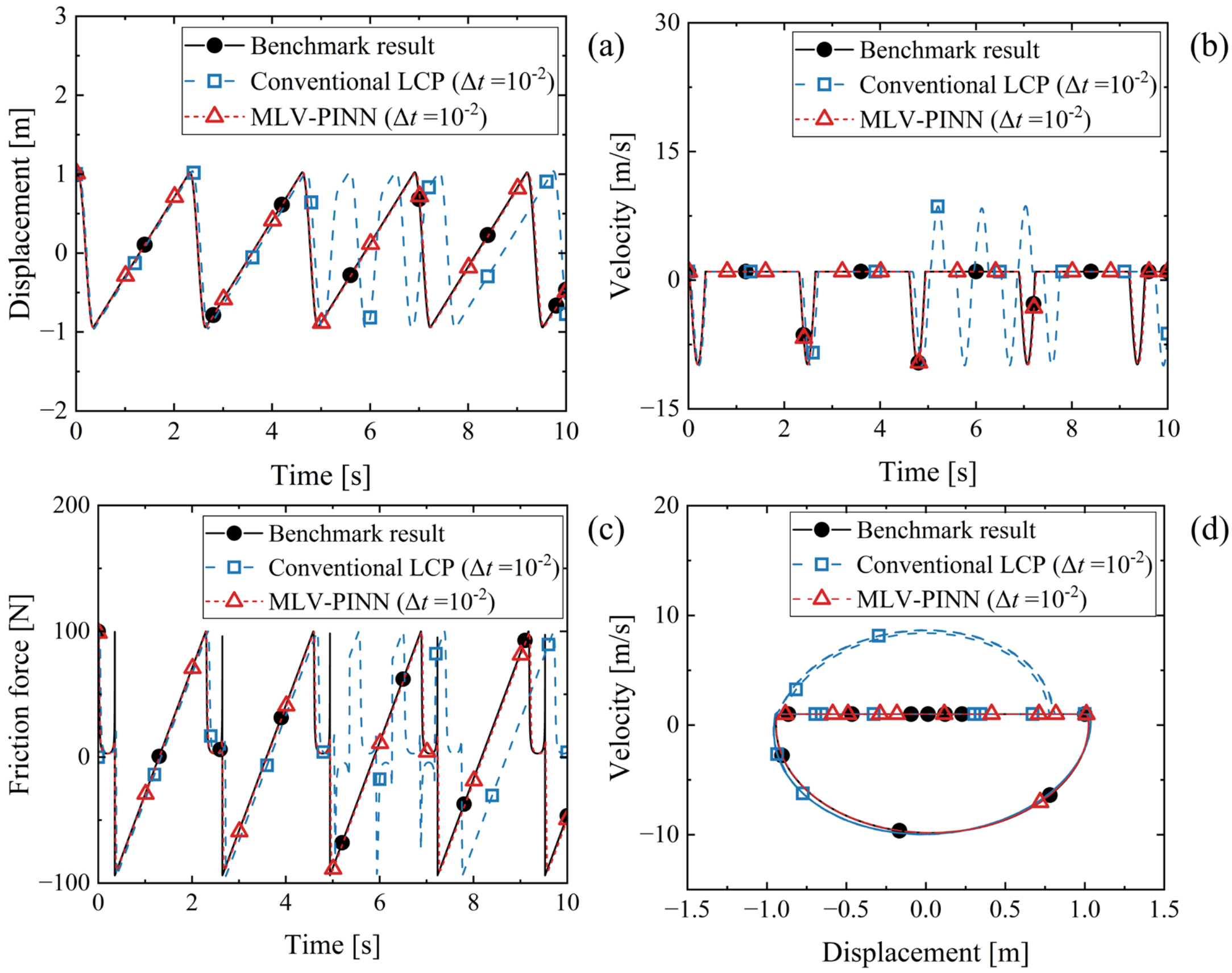

Fig. 6. Comparison of stick-slip vibration responses in Case I: (a) displacement time history; (b) velocity time history; (c) friction-force time history; (d) phase trajectory.

### *3.3. Case II: One-dimensional nonsmooth simulation with coupled stick-slip at multiple contact points*

#### *3.3.1. Physical problem and formulation*

Building on Case I, this section considers a system of two sliders connected in series on a moving belt to assess MLV-PINN for coupled vibration with tangential nonsmoothness at multiple contact points. Each slider can undergo stick-slip transitions at its contact with the belt, while the connecting spring couples their motion. An error in identifying the transition at either contact point may affect the overall dynamic response.

As shown in Fig. 7(a), the upper part of the system consists of two sliders with masses $m_1$ and $m_2$, both in frictional contact with the belt below. The belt moves at a constant velocity $v_0$. Slider $m_1$ is connected to a fixed support through spring $k_1$, while sliders $m_1$ and $m_2$ are coupled by spring $k_2$. The normal loads acting on the two sliders are denoted by $N_1$ and $N_2$, respectively. Each contact point may be in either the sticking or sliding state, giving four possible state combinations and transitions between them, as illustrated in Fig. 7(b).

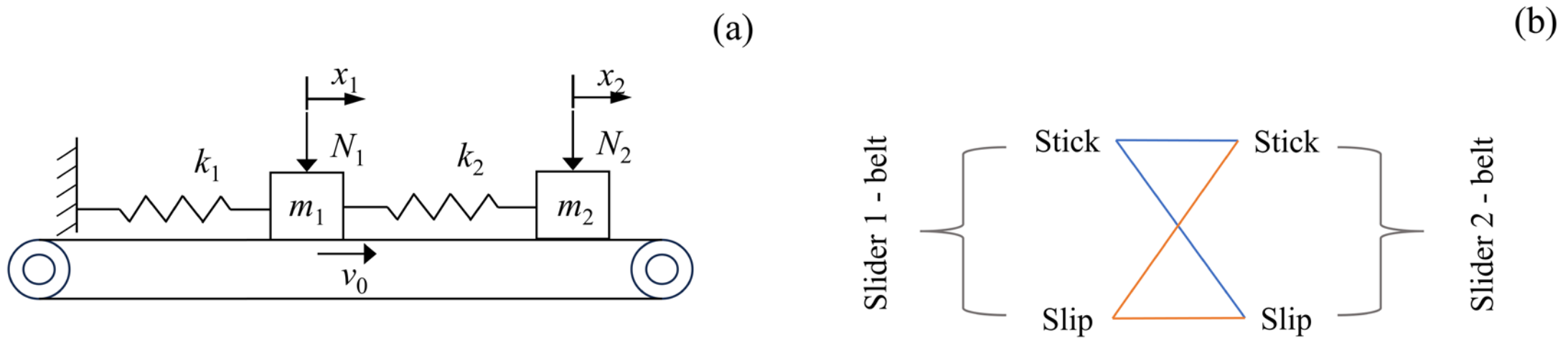

Fig. 7. Mechanical model of Case II: (a) multi-slider-belt system; (b) combinations of sticking and sliding states.

In this system, the independent coordinates are the absolute displacements $x_1$ and $x_2$ of the two sliders along the direction of belt motion. With the displacement vector defined as $\mathbf{x}(t) = [x_1(t), x_2(t)]^{\mathrm{T}}$, the equations of motion can be written as

$$\mathbf{M}\ddot{\mathbf{x}} + \mathbf{K}\mathbf{x} - \mathbf{W}_{\mathrm{T}}\boldsymbol{\lambda}_{\mathrm{T}} = 0 \tag{19}$$

where the mass matrix $\mathbf{M}$, stiffness matrix $\mathbf{K}$, tangential generalised force direction matrix $\mathbf{W}_{\mathrm{T}}$, and friction-force vector $\boldsymbol{\lambda}_{\mathrm{T}}$ are given by

$$\mathbf{M} = \begin{bmatrix} m_1 & 0 \\ 0 & m_2 \end{bmatrix}, \mathbf{K} = \begin{bmatrix} k_1 + k_2 & -k_2 \\ -k_2 & k_2 \end{bmatrix}, \mathbf{W}_{\mathrm{T}} = \begin{bmatrix} 1 & 0 \\ 0 & 1 \end{bmatrix}, \boldsymbol{\lambda}_{\mathrm{T}} = \begin{bmatrix} \lambda_{\mathrm{T}1} \\ \lambda_{\mathrm{T}2} \end{bmatrix} \tag{20}$$

In MLV-PINN, the residual of Eq. (19) is evaluated at each IRK stage to construct the coupled equations-of-motion residual loss $\mathcal{L}_{\mathrm{dyn}}$. This loss is combined with the tangential MLV complementary loss $\mathcal{L}_{\mathrm{T}}$ in Eq. (13) and the friction-margin consistency loss $\mathcal{L}_{\mathrm{cons}}$ in Eq. (14), evaluated at both contact points, to form the composite physics-informed loss for the current time step.

*3.3.2. Results and discussion*

Using the same benchmark result procedure and time-step sizes as in Case I, MLV-PINN is compared with the benchmark result and the conventional LCP method. The model parameters are $m_1 = m_2 = 1.0$, $k_1 = k_2 = 10.0$, $v_0 = 1.0$, $N_1 = 60$, and $N_2 = 40$. The friction model is given by Eq. (16), with $\mu_0 = 0.4$ and $\delta = 10$. These parameters allow stick-slip transitions at both contact points. The initial conditions are $x_1(0) = 10.0$, $\dot{x}_1(0) = 1.0$, $x_2(0) = 10.0$, and $\dot{x}_2(0) = 1.0$.

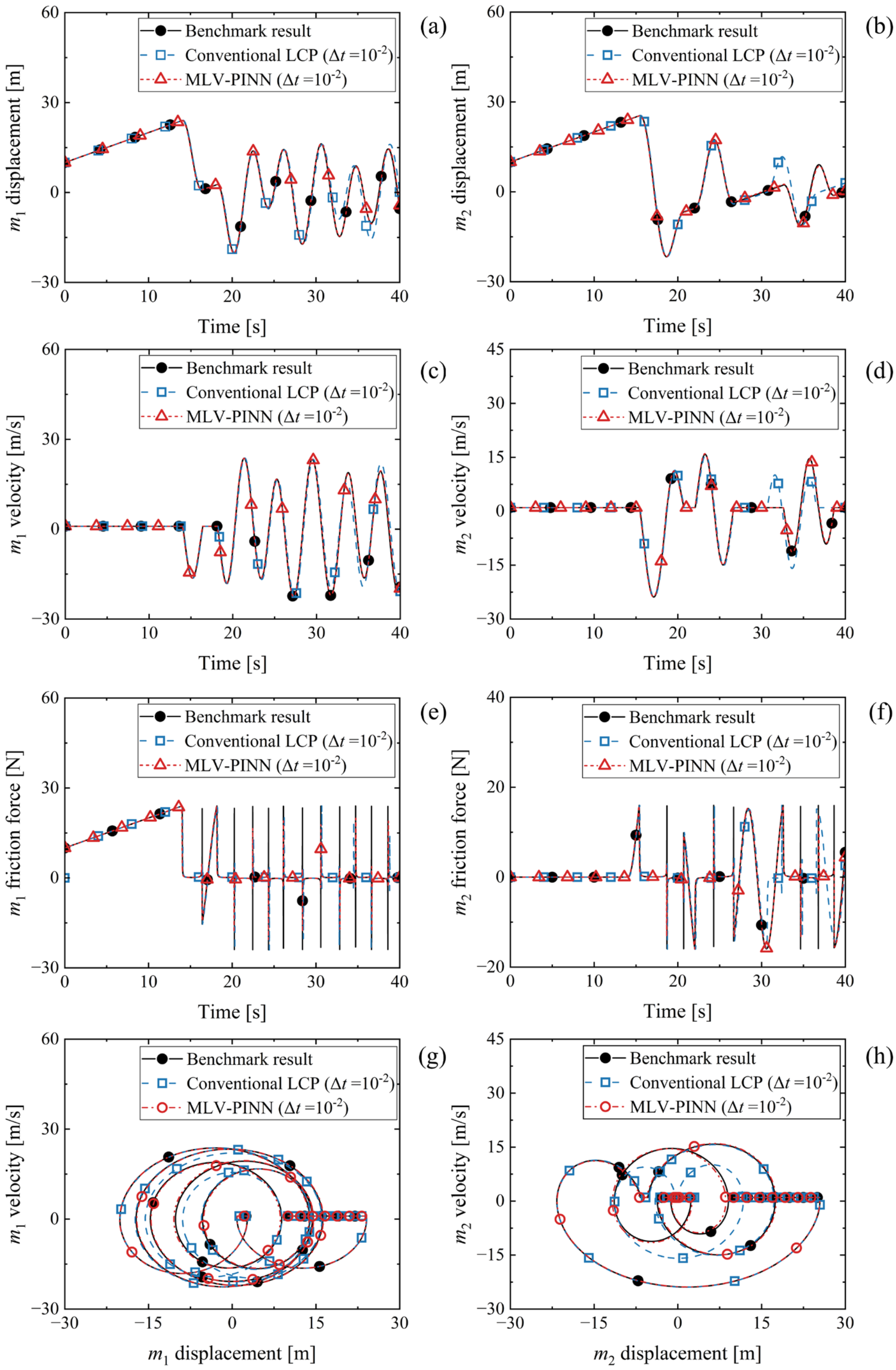

Fig. 8. Comparison of coupled stick-slip vibration responses in Case II: (a) displacement time history of slider 1; (b) displacement time history of slider 2; (c) velocity time history of slider 1; (d) velocity time history of slider 2; (e) friction-force time history of slider 1; (f) friction-force time history of slider 2; (g) phase trajectory of slider 1; (h) phase trajectory of slider 2.

Figs. 8(a)-8(h) present the displacement, velocity, and friction-force time histories and the displacement-velocity phase trajectories of the two sliders. The curve colours and markers are consistent with those used in Case I. As shown in Figs. 8(a)-(b), the displacement responses predicted by MLV-PINN for both sliders remain in close agreement with the benchmark result, whereas those predicted by the conventional LCP method gradually deviate over 30-40 s.

Figs. 8(c)-(d) show the velocity responses of the two sliders. The overall dynamic-response error produced by the conventional LCP method during 30-40 s arises from an erroneous stick-slip transition at the contact point of slider $m_2$. As confirmed by the friction-force response in Fig. 8(f), this transition error leads to incorrect switching of the friction branch and, through dynamic coupling, causes deviations in the velocity responses of both sliders. By contrast, MLV-PINN maintains velocity predictions closer to the benchmark result for both degrees of freedom. The friction-force response in Fig. 8(f) also shows that the conventional LCP method exhibits incorrect switching of the friction branch associated with the state-transition error, whereas the friction forces predicted by MLV-PINN at both contact points remain closer to the benchmark result. The phase trajectories in Figs. 8(g)-(h) also show that MLV-PINN preserves the connection between the sticking segments and the sliding branches.

### *3.4. Case III: Two-dimensional nonsmooth simulation with coupled stick-slip and contact-separation*

#### *3.4.1. Physical problem and formulation*

Case III extends the analysis to a more challenging two-dimensional problem involving coupled transverse and longitudinal motion, with nonsmoothness in both the normal and tangential directions. To further examine the performance of the proposed method under these coupled nonsmooth effects, this section considers the frictional system with coupled transverse and longitudinal motion shown in Fig. 9(a). This model describes friction-induced vibration caused by coupling between friction and structural modes. The resulting mode-coupling instability can induce separation-reattachment transitions at the interface, which introduces normal nonsmoothness.

In the model shown in Fig. 9(a), a slider of mass $m$ is placed on a belt moving at a constant velocity $v_0$. The slider is constrained in the tangential direction by a spring of stiffness $k_1$ and is also connected to an inclined spring of stiffness $k_2$ oriented at 45° to the horizontal. The inclined spring $k_2$ couples the horizontal and normal motions, so the tangential friction force can affect both the horizontal response and the normal contact state. Fig. 9(b) illustrates the possible contact states of the slider, including normal contact-separation and tangential stick-slip transitions.

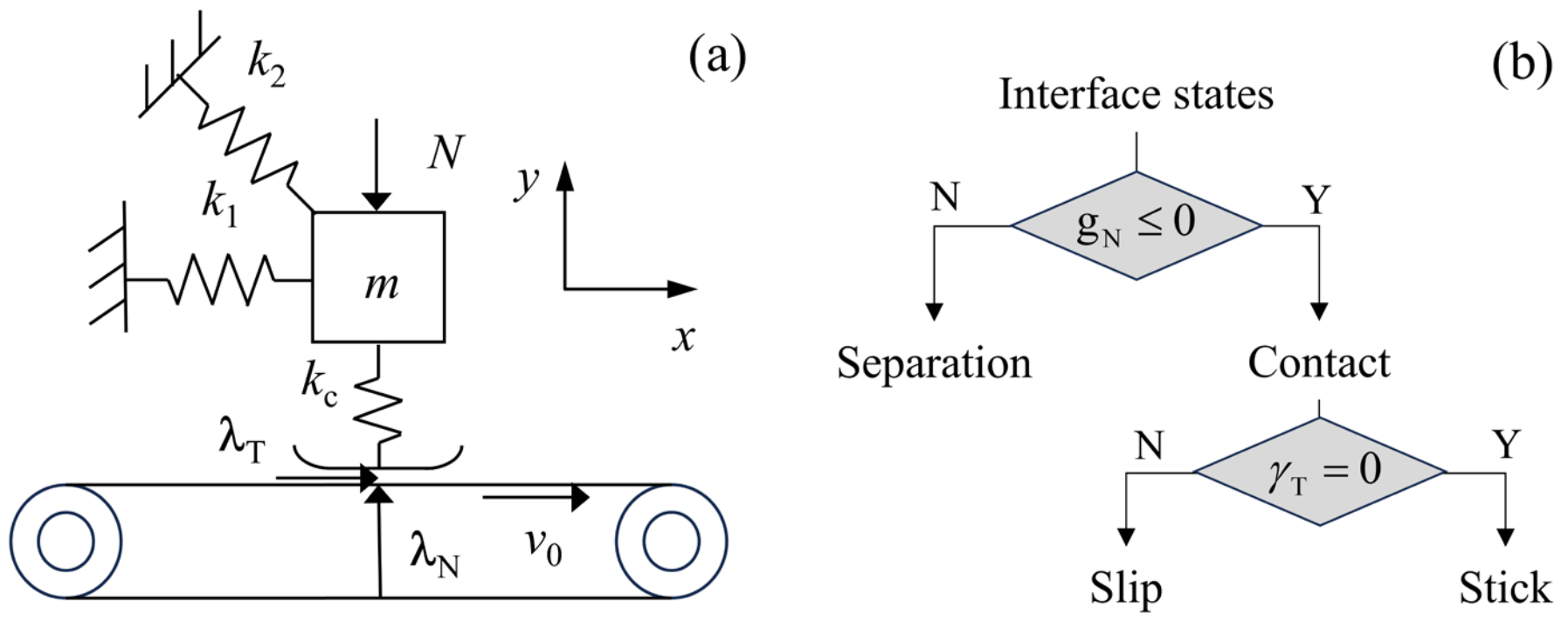

Fig. 9. Mechanical model of Case III: a two-degree-of-freedom frictional vibration system with nonsmooth normal contact and tangential friction: (a) model schematic; (b) contact states of the slider.

Define the generalised coordinate vector as $\mathbf{q} = [x, y]^{\mathrm{T}}$, where $x$ and $y$ denote the absolute displacements of the slider in the horizontal and vertical directions, respectively. The equations of motion are then written as

$$\mathbf{M}\ddot{\mathbf{q}} + \mathbf{K}\mathbf{q} - \mathbf{W}_{\mathrm{N}}\boldsymbol{\lambda}_{\mathrm{N}} - \mathbf{W}_{\mathrm{T}}\boldsymbol{\lambda}_{\mathrm{T}} = \mathbf{F}_{\mathrm{e}} \tag{21}$$

where the mass matrix $\mathbf{M}$, stiffness matrix $\mathbf{K}$, external force vector $\mathbf{F}_{\mathrm{e}}$, normal generalised force direction matrix $\mathbf{W}_{\mathrm{N}}$, and tangential generalised force direction matrix $\mathbf{W}_{\mathrm{T}}$ are given by

$$\mathbf{M} = \begin{bmatrix} m & 0 \\ 0 & m \end{bmatrix}, \mathbf{K} = \begin{bmatrix} k_1 + k_2/2 & -k_2/2 \\ -k_2/2 & k_2/2 \end{bmatrix}, \mathbf{F}_{\mathrm{e}} = \begin{bmatrix} 0 \\ N \end{bmatrix}, \mathbf{W}_{\mathrm{N}} = \begin{bmatrix} 0 \\ 1 \end{bmatrix}, \mathbf{W}_{\mathrm{T}} = \begin{bmatrix} 1 \\ 0 \end{bmatrix} \tag{22}$$

In MLV-PINN, the residual of Eq. (21) is evaluated at each IRK stage to construct the coupled equations-of-motion residual loss $\mathcal{L}_{\mathrm{dyn}}$. This loss is combined with the normal-contact complementary loss $\mathcal{L}_{\mathrm{N}}$, tangential MLV complementary loss $\mathcal{L}_{\mathrm{T}}$, and friction-margin consistency loss $\mathcal{L}_{\mathrm{cons}}$ to form the composite physics-informed loss for the current time step.

### *3.4.2. Results and discussion*

The Switch Model used as the benchmark in the preceding cases is no longer applicable as this case involves both normal contact-separation and tangential stick-slip transitions. The benchmark result is therefore obtained using a pseudo-RK4 time-stepping method (hereafter referred to as RK4-LCP) with a small time step of $\Delta t = 10^{-4}$. For comparison, MLV-PINN, the conventional LCP method, and the RK4-LCP method all use $\Delta t = 10^{-2}$. The model parameters are $m = 5.0$, $k_1 = 500$, $k_2 = 300$, $k_c = 250$, $v_0 = 2.0$, and $N = 400$. The friction model is given by Eq. (17), with $\mu_k = 0.1$, $\mu_s = 0.2$, and $\alpha = 3$. Owing to the coupled horizontal-vertical motion in this case, the relative sliding velocity spans a relatively wide range. If the rational friction model were used, the friction coefficient could decay to an excessively low level at higher sliding speeds, lowering the friction limit and shortening the observable sticking intervals. Therefore, the exponential model is adopted. The initial conditions are $x(0) = -10.0$, $y(0) = -10.0$, $\dot{x}(0) = 2.0$, and $\dot{y}(0) = 0.0$.

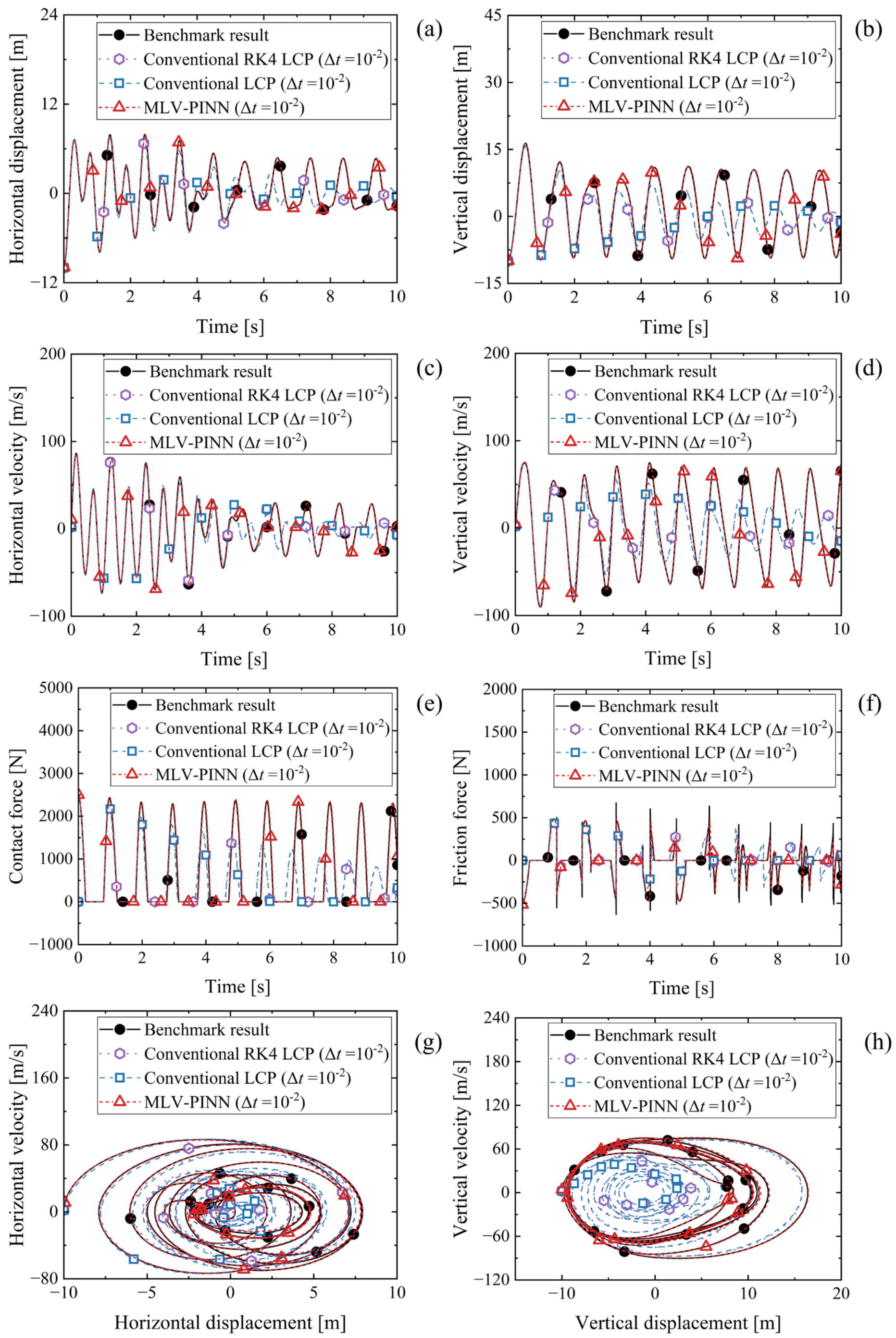

Fig. 10. Comparison of contact-separation and stick-slip vibration responses in Case III: (a) horizontal displacement time history; (b) normal displacement time history; (c) horizontal velocity time history; (d) normal velocity time history; (e) contact-force time history; (f) friction-force time history; (g) phase trajectory of horizontal vibration; (h) phase trajectory of normal vibration.

Figs. 10(a)-10(h) present the time histories of the horizontal and normal displacements, horizontal and normal velocities, normal contact force and friction force, together with the corresponding displacement-velocity phase planes, obtained by the three methods and the benchmark result. The black solid line with filled-circle markers denotes the benchmark result, the purple dotted line with hexagonal markers denotes the RK4-LCP method, the blue dashed line with rectangular markers denotes the conventional LCP method, and the red short-dashed line with upward-pointing triangular markers denotes MLV-PINN. This case tests whether the methods can capture both normal contact-separation and tangential stick-slip transitions.

As shown in Figs. 10(a)-(b), both the conventional first-order LCP method and the RK4-LCP method exhibit errors in the horizontal and vertical vibration responses of the slider, with errors appearing first in the vertical direction. The velocity responses in Figs. 10(c)-(d) and the contact-force responses in Figs. 10(e)-(f) indicate that these errors originate from inaccuracies in the normal contact force emerging at $t = 2$ s. These errors subsequently propagate to the tangential motion, causing the predicted stick-slip transitions to deviate significantly from the benchmark result. The resulting relative-velocity error also shifts the friction-branch switching time. By comparison, MLV-PINN predicts both contact-separation and stick-slip transitions more closely to the benchmark result. The phase portraits in Figs. 10(g)-(h) further show that MLV-PINN preserves the phase-trajectory structure of the horizontal and normal motions, whereas the conventional methods exhibit clear trajectory deviations near state transitions. These results demonstrate that MLV-PINN can capture the main nonsmooth state transitions with good accuracy even at a relatively large time step in a coupled system involving both normal contact-separation and tangential stick-slip.

### *3.5. Case IV: Nonsmooth simulation of a continuum system with friction*

#### *3.5.1. Physical problem and formulation*

Most existing numerical methods for nonsmooth dynamics deal with systems with a finite number of degrees of freedom and multiple contact points. The first three cases also belong to this class of discrete systems. To further examine the applicability of MLV-PINN to continuum systems, this section considers an elastic beam in frictional contact with a moving slider and uses the proposed method to calculate its coupled normal-tangential nonsmooth dynamic response.

Fig. 11(a) illustrates the infinite DoF model with a continuum beam and a discrete body,

representative of the frictional system with moving-load. In such engineering problems, both the upper and lower subsystems participate in the vibration. The lower contact body is modelled as a simply supported Euler-Bernoulli beam of span $L$. The upper subsystem is a slider driven by a horizontal spring-damper element. One end is connected to the slider, and the other moves to the right at a constant velocity $v_1$. The slider has mass $m$ and initial horizontal position $x_0$. The horizontal spring has stiffness $k_1$ and is initially undeformed. The top of the slider is connected to a normal spring-damper element of stiffness $k_2$ and is subjected to a constant normal load $N$. The normal spring-damper element moves along the beam together with the slider and is assumed to provide spring and damping forces only in the normal direction. The model includes normal separation-reattachment transitions, together with tangential stick-slip transitions between the slider and the beam. Compared with the preceding lumped-parameter models, it also accounts for a moving contact location and the multimodal response of the beam. The corresponding contact states are illustrated in Fig. 11(b). The normal contact state is determined by the normal gap between the beam and the slider, which is coupled with horizontal stick-slip motion of the slider.

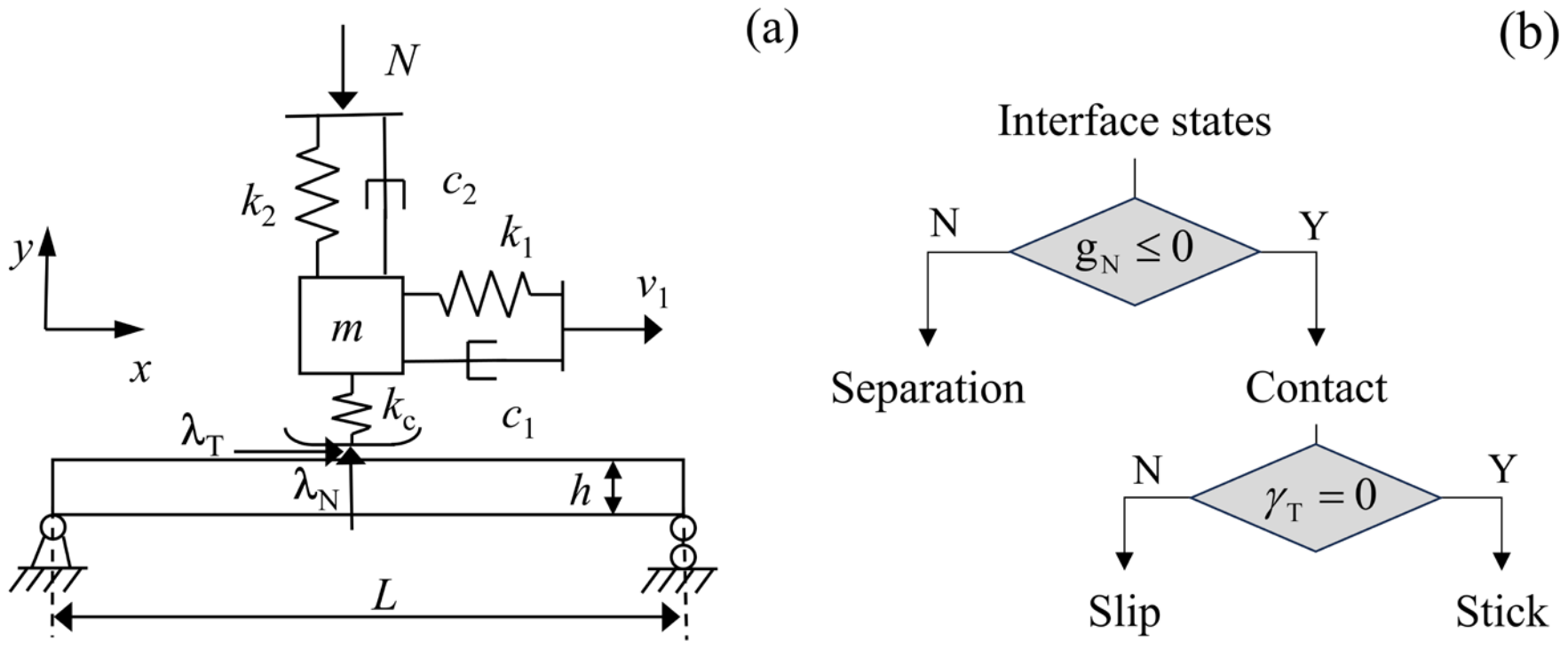

Fig. 11. Mechanical model of Case IV: an elastic beam-slider system with moving contact and coupled normal-tangential nonsmoothness: (a) model schematic; (b) beam-slider contact states.

The absolute displacements of the slider in the horizontal and vertical directions are denoted by $x(t)$ and $y(t)$, respectively. Let $x_\mathrm{b}$ denote the spatial coordinate of the simply supported beam, and $w(x_\mathrm{b},t)$ its transverse displacement field. As the contact location moves with the slider, the contact force acting on the beam is represented as a moving concentrated load. The governing equation for the transverse vibration of the beam can be written as

$$\rho A\ddot{w}(x_\mathrm{b},t)+c_\mathrm{b}\dot{w}(x_\mathrm{b},t)+EIw''''(x_\mathrm{b},t)=\lambda_\mathrm{N}(t)\delta(x_\mathrm{b}-x_\mathrm{c}(t)) \tag{23}$$

where $\rho A$ is the mass per unit length, $EI$ denotes the flexural rigidity, $c_\mathrm{b}$ is the equivalent structural damping coefficient, $\lambda_\mathrm{N}(t)$ is the normal contact force, and $\delta$ denotes the Dirac delta function. Here, $x_\mathrm{c}(t)$ denotes the instantaneous horizontal position of the beam-slider contact point. The simply supported boundary conditions are given by:

$$w(0,t)=w(L,t)=0, w''(0,t)=w''(L,t)=0 \tag{24}$$

The moving load in Eq. (23) introduces both spatial localisation and explicit time dependence into the external forcing. If the network directly predicted $w(x_{\mathrm{b}},t)$, it would need to approximate both the smooth global beam response and the local effects induced by the moving concentrated load, which makes training difficult. To avoid treating the Dirac function directly, modal expansion is used [45] to project the continuum equation onto a finite-dimensional modal space. This transforms the partial differential equation into a set of second-order ordinary differential equations in generalised coordinates.

The beam displacement field is approximated by a truncated expansion of the first $N_m$ modes:

$$w(x_{\mathrm{b}},t)=\sum_{r=1}^{N_m}\phi_r(x_{\mathrm{b}})\eta_r(t) \tag{25}$$

Here, $\eta_r(t)$ denotes the generalised coordinate of the $r$th mode. For a simply supported beam, the spatial mode shapes are expressed as:

$$\phi_r(x_{\mathrm{b}})=\sqrt{\frac{2}{\rho AL}}\sin\frac{r\pi x_{\mathrm{b}}}{L}, r=1,2,\cdots,N_m. \tag{26}$$

These mode shapes satisfy the orthonormality condition $\int_0^L \rho A\phi_r(x_{\mathrm{b}})\phi_s(x_{\mathrm{b}})\mathrm{d}x=\delta_{rs}$. Substituting Eq. (25) into Eq. (23) and applying Galerkin projection yields the equations of motion of the beam in modal coordinates. The equation for the $r$th mode is

$$\ddot{\eta}_r+2\xi_r\omega_r\dot{\eta}_r+\omega_r^2\eta_r=-\lambda_{\mathrm{N}}(t)\phi_r(x_{\mathrm{c}}(t)), \quad (r=1,2,\ldots,N_m) \tag{27}$$

where $\xi_r$ and $\omega_r$ denote the damping ratio and natural frequency of the $r$th beam mode, respectively. After the modal projection, the Dirac function no longer appears explicitly in the governing equations, and the effect of the moving contact force on the beam is represented by the generalised modal force $\lambda_{\mathrm{N}}(t)\phi_r(x_{\mathrm{c}}(t))$.

The equations of motion of the slider can be written as:

$$\mathbf{M}\ddot{\mathbf{q}}+\mathbf{C}\dot{\mathbf{q}}+\mathbf{K}\mathbf{q}-\mathbf{W}_{\mathrm{N}}\boldsymbol{\lambda}_{\mathrm{N}}-\mathbf{W}_{\mathrm{T}}\boldsymbol{\lambda}_{\mathrm{T}}=\mathbf{F}_{\mathrm{e}} \tag{28}$$

where $\mathbf{M}=\begin{bmatrix} m & 0 \\ 0 & m \end{bmatrix}$, $\mathbf{C}=\begin{bmatrix} c_1 & 0 \\ 0 & c_2 \end{bmatrix}$, $\mathbf{K}=\begin{bmatrix} k_1 & 0 \\ 0 & k_2 \end{bmatrix}$, $\mathbf{F}_{\mathrm{e}}(t)=\begin{bmatrix} k_1(x_0+v_1t)+c_1v_1 \\ -\mathrm{N} \end{bmatrix}$, $\mathbf{W}_{\mathrm{N}}=\begin{bmatrix} 0 \\ 1 \end{bmatrix}$, $\mathbf{W}_{\mathrm{T}}=\begin{bmatrix} 1 \\ 0 \end{bmatrix}$.

According to Eq. (25), the transverse displacement of the beam at the contact point is:

$$w_{c}(t)=w(x_{c}(t),t)=\sum_{r=1}^{N_m}\eta_r(t)\phi_r(x_{c}(t)) \tag{29}$$

Accordingly, the normal gap between the slider and the beam can be written as:

$$g_{N}(t)=y(t)-w_{c}(t)-h_0=y(t)-\sum_{r=1}^{N_m}\phi_r(x_{c}(t))\eta_r(t)-h_0 \tag{30}$$

where $h_0$ denotes the initial geometric gap or the reference distance for contact.

In the tangential direction, the relative velocity is defined as:

$$\gamma_{T}(t)=\dot{x}(t) \tag{31}$$

The tangential friction force is expressed as:

$$\lambda_{T}(t)=\frac{\lambda_{R}(t)-\lambda_{L}(t)}{2} \tag{32}$$

where $\lambda_{R}$ and $\lambda_{L}$ are the two non-negative friction margins, and the friction bounds are jointly determined by the velocity-dependent friction coefficient and the normal contact force. The tangential friction constraint is described by the MLV-based complementary relations in Eq. (8) and enforced through the tangential complementary loss $\mathcal{L}_{T}$ in Eq. (13). The friction-margin consistency loss $\mathcal{L}_{cons}$ is given by Eq. (14). The normal contact force is determined by the unilateral-contact complementary relation in Eq. (3), which is imposed through the loss $\mathcal{L}_{N}$ in Eq. (12). Together with the losses constructed from the beam modal equations in Eq. (27) and the slider equations of motion in Eq. (28), these terms form the composite physics-informed loss function for the present case.

MLV-PINN can therefore be extended to the nonsmooth dynamic analysis of this continuum system. The network inputs are the initial states of the slider and the beam at the beginning of the current time step, including the slider displacements $(x,y)$, slider velocities $(\dot{x},\dot{y})$, modal displacements $\{\eta_r\}_{r=1}^{N_m}$, and modal velocities $\{\dot{\eta}_r\}_{r=1}^{N_m}$. The network outputs are the beam modal accelerations $\{\ddot{\eta}_r^{(i)}\}_{r=1}^{N_m}$, slider accelerations $\mathbf{a}^{(i)}$, normal contact force $\boldsymbol{\lambda}_{N}^{(i)}$, and positive- and negative-direction tangential friction margins $\boldsymbol{\lambda}_{R}^{(i)}$ and $\boldsymbol{\lambda}_{L}^{(i)}$ at all IRK stages. The stage states are reconstructed from the beam and slider equations of motion in Eqs. (27)-(28), while the contact kinematics and friction quantities are evaluated from Eqs. (29)-(32). Together with the normal and tangential complementary relations in Eqs. (3) and (8), these relations form the physical basis for the loss evaluation in Step 4 of Fig. 4.

*3.5.2. Results and discussion*

Because the contact location varies with time and the normal gap depends on the beam displacement at the moving contact point, a fixed contact mapping is not readily applicable in a conventional LCP-based time-stepping formulation. An event-driven method is therefore used to monitor state transitions and locate separation-reattachment and stick-slip events by bisection to obtain the benchmark result [46]. The benchmark result is computed by combining the event-driven scheme with fourth-order Runge-Kutta integration using a small time step of $\Delta t = 10^{-4}$. For comparison, the time step of the event-driven method is increased to $\Delta t = 10^{-2}$, while MLV-PINN also uses $\Delta t = 10^{-2}$.

The model parameters are $L = 2$, $EI = 10$, $\rho A = 1$, $c_b = 0.01$, $m = 2.0$, $k_1 = 10$, $k_2 = 200$, $c_1 = 0.05$, $c_2 = 0.01$, $v_1 = 0.03$, $N = 2.9$, $N_m = 5$, and $k_c = 10^5$. The friction model is given by Eq. (17), with $\mu_s = 0.6$, $\mu_k = 0.2$, and $\alpha = 50$. The initial conditions for the slider are $x(0) = 0.1$, $y(0) = 0.0$, $\dot{x}(0) = 0.0$, and $\dot{y}(0) = 0.0$. For the beam, the initial conditions for the first mode are $\eta_1(0) = 0.0$ and $\dot{\eta}_1(0) = -1.0$, while the initial modal displacements and velocities of all remaining modes are set to zero.

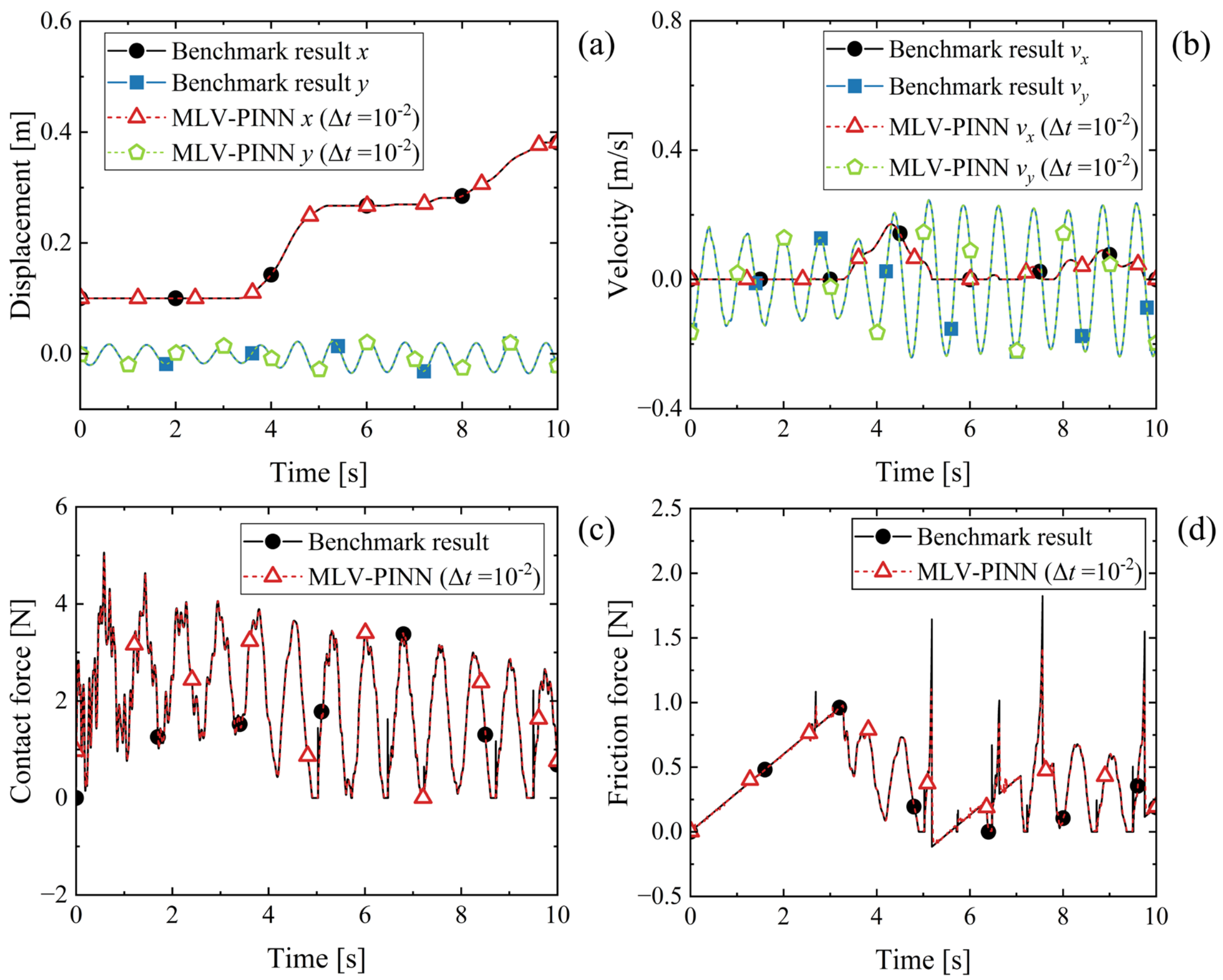

Fig. 12. Comparison of slider contact-separation and stick-slip vibration responses in Case IV: (a) displacement time histories; (b) velocity time histories; (c) normal contact-force time history; (d) friction-force time history.

Figs. 12(a)-(d) present the time histories of the slider displacement, velocity, normal contact force, and tangential friction force. The black solid lines with filled-circle markers and the blue solid lines with filled-rectangle markers denote the benchmark responses in the $x$ and $y$ directions, respectively, while the red short-dashed lines with upward-pointing triangular markers and the green short-dashed lines with pentagonal markers denote the corresponding MLV-PINN predictions. As shown in Figs. 12(a) and 12(b), MLV-PINN results closely follow the displacement and velocity responses of the slider in both directions without noticeable phase drift. Figs. 12(c)-(d) further show that MLV-PINN accurately captures the variations in the normal contact force and the switching of the tangential friction branch. At the same relatively large time step, the event-driven method exhibits pronounced numerical instability and progressive error growth due to accumulated event-localisation errors; the corresponding results are provided in Fig. S1 of the Supplementary Material. These results show that MLV-PINN provides stable predictions of the nonsmooth response when the moving contact is coupled with the beam modal response.

### 3.6. Error analysis of different methods in the four cases

To compare the accuracy of MLV-PINN and conventional numerical methods at relatively large time steps, Table 1 summarises the main RMSE measures for the four cases. The reported errors are those in displacement, velocity, friction force, and phase-plane response. For Cases III and IV, which involve normal contact-separation, the normal contact-force RMSE is also reported. For multi-degree-of-freedom cases, the displacement and velocity RMSEs are obtained by averaging the corresponding RMSEs over all degrees of freedom. For each displacement-velocity pair, the composite phase-plane error at each time instant is defined as $\sqrt{(\Delta q)^2 + (\Delta v)^2}$, where $\Delta q$ and $\Delta v$ denote the displacement and velocity deviations between the numerical result and the benchmark result, respectively. The corresponding RMSE is then averaged over all degrees of freedom. All RMSE values are evaluated on a common temporal grid with $\Delta t = 10^{-2}$, with the high-fidelity benchmark result interpolated to the corresponding sampling instants.

Table 1. RMSE comparison of MLV-PINN and conventional methods at relatively large time steps in the four cases ($\Delta t = 10^{-2}$).

| Case | Comparison method | Disp. RMSE | Vel. RMSE | Fric. RMSE | Cont. RMSE | Phase RMSE |
|---|---|---|---|---|---|---|
| I | Conventional LCP | $7.1\times10^{-1}$ | 4.3 | $5.7\times10^{1}$ | — | 4.3 |

| Case | Comparison method | Disp. RMSE | Vel. RMSE | Fric. RMSE | Cont. RMSE | Phase RMSE |
|---|---|---|---|---|---|---|
| | MLV-PINN | $3.5\times10^{-2}$ | $3.4\times10^{-1}$ | $1.0\times10^{1}$ | — | $3.5\times10^{-1}$ |
| II | Conventional LCP | 2.3 | 2.3 | 2.9 | — | 3.3 |
| | MLV-PINN | $2.9\times10^{-1}$ | $3.4\times10^{-1}$ | 1.3 | — | $4.5\times10^{-1}$ |
| III | Conventional RK4-LCP | 3.8 | $2.6\times10^{1}$ | $1.2\times10^{2}$ | $7.7\times10^{2}$ | $2.6\times10^{1}$ |
| | MLV-PINN | $2.0\times10^{-1}$ | 1.7 | $5.9\times10^{1}$ | $8.5\times10^{1}$ | 1.7 |
| IV | Event-driven method | $9.1\times10^{-2}$ | $6.6\times10^{-1}$ | 7.8 | $3.9\times10^{1}$ | $6.7\times10^{-1}$ |
| | MLV-PINN | $1.2\times10^{-4}$ | $3.2\times10^{-4}$ | $4.0\times10^{-2}$ | $9.6\times10^{-2}$ | $3.5\times10^{-4}$ |

As shown in Table 1, MLV-PINN yields lower RMSE values than the corresponding conventional methods at relatively large time steps for the principal response quantities in all four cases. For Case I, the displacement, velocity, and phase-plane RMSEs decrease from $7.1\times10^{-1}$, $4.3$, and $4.3$ to $3.5\times10^{-2}$, $3.4\times10^{-1}$, and $3.5\times10^{-1}$, respectively. This indicates that, for the single-degree-of-freedom stick-slip system, MLV-PINN reduces the response deviations associated with errors in stick-slip transition identification. For Case II, the displacement, velocity, and phase-plane errors averaged over the two sliders are also reduced, showing that MLV-PINN remains closer to the benchmark result when asynchronous stick-slip transitions occur in the coupled system.

Case III involves both normal contact-separation and tangential stick-slip transitions, with errors arising from the normal-contact timing, tangential friction-branch switching, and horizontal-vertical coupling. Table 1 shows that MLV-PINN gives lower RMSE values for displacement, velocity, contact force, and phase-plane response than the corresponding conventional method at the same relatively large time step. For example, the contact-force RMSE decreases from $7.7\times10^{2}$ to $8.5\times10^{1}$, while the phase-plane RMSE decreases from $2.6\times10^{1}$ to $1.7$. The relative reduction in friction-force RMSE for Cases I-III is smaller than those of the displacement and velocity metrics. This is associated with the sharp local variations in the friction-force time histories. For such nonsmooth responses, even small discrepancies in the switching time or peak amplitude can produce relatively large RMSE values. The friction-force error should therefore be interpreted together with the corresponding time histories. As shown in Figs. 6(c), 8(e)-8(f), and 10(f), MLV-PINN remains in closer agreement with the benchmark result in the timing of friction-branch switching and abrupt force changes.

Fig. 13 presents the error-reduction ratios of MLV-PINN relative to the conventional methods at relatively large time steps across the four cases. The error-reduction ratio is defined as

$$R_{\mathrm{imp}} = \frac{E_{\mathrm{conv}} - E_{\mathrm{MLV}}}{E_{\mathrm{conv}}} \times 100\% \tag{33}$$

where $E_{conv}$ and $E_{MLV}$ denote the RMSE values of the conventional numerical method with a relatively large time step and MLV-PINN, respectively. The cases differ in their physical quantities, response amplitudes, and nonsmooth mechanisms. The ratios in Fig. 13 therefore compare the two methods for the same metric within each case. The absolute errors are given by the RMSE values reported in Table 1.

As shown in Fig. 13, MLV-PINN reduces the displacement, velocity, and phase-plane RMSEs by 85.16%-99.95% relative to the conventional methods at relatively large time steps in the four cases. The reductions in friction-force RMSE are smaller in Cases II and III, at 53.20% and 53.12%, respectively. Case IV shows the largest improvements, with reductions of 99.48%-99.95% in all five metrics. These results support the accuracy improvements observed in the preceding time-history and phase-plane comparisons, including the extension to the beam-slider continuum system.

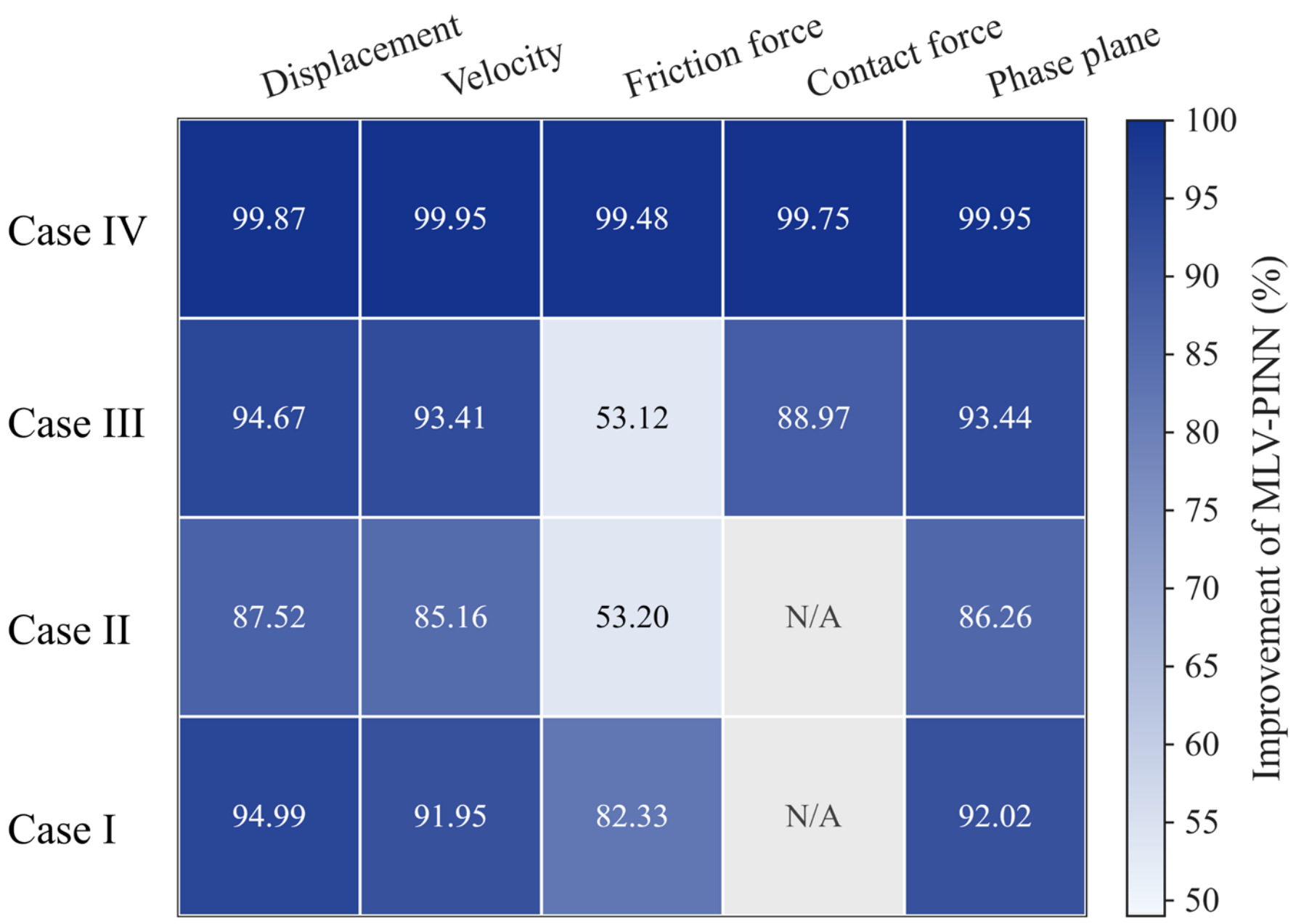

Fig. 13. Percentage reductions in RMSE achieved by MLV-PINN.

## 4. Factors affecting the performance of MLV-PINN

This section examines four factors affecting the performance of MLV-PINN: the tangential leading variable, the friction-velocity slope, the simulation duration, and the training configuration. The comparisons focus on state-transition predictions, response errors, and total-loss evolution.

### *4.1. Effect of the tangential leading variable*

This section investigates the role of the mixed leading variable in identifying sticking and sliding states through an ablation study. The three sets of calculations differ only in the leading variable used in the tangential complementary relation and use the same network architecture, time step, optimisation strategy, and loss weights. For the single-degree-of-freedom system in Case I, predictions from a relative-velocity-based PINN, an acceleration-based PINN, and MLV-PINN are compared with the Switch Model benchmark result. The acceleration-based PINN is used here solely as an alternative residual formulation for the ablation study and should not be interpreted as a conventional acceleration-level frictional complementary solver.

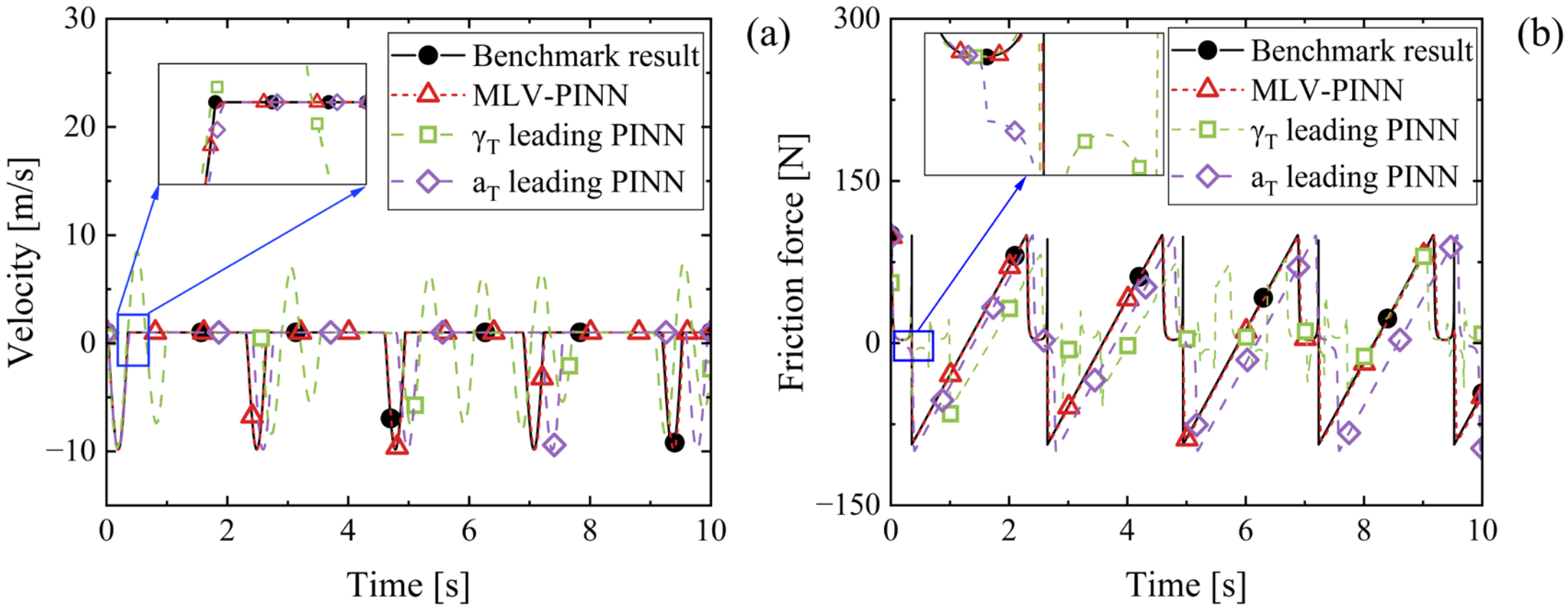

Fig. 14. Ablation study of the tangential leading variable: (a) velocity time history; (b) friction-force time history.

Figs. 14(a)-(b) compare the velocity and friction-force time histories predicted by the three PINN models. The relative-velocity-based PINN exhibits delayed state transitions and velocity overshoot as the relative velocity approaches zero. The acceleration-based PINN exhibits spurious friction-branch switching, causing the friction force to deviate from the benchmark result. MLV-PINN agrees more closely with the benchmark result near stick-slip transitions, predicting the transition times and associated friction-force changes more accurately.

These differences reflect the information provided by each leading variable. The relative velocity identifies the current sliding direction, but its instantaneous value alone provides limited information about an impending transition within the current time step. Acceleration describes the change in velocity and may change sign even when the sliding direction remains unchanged. Using acceleration alone may therefore cause incorrect friction-branch switching. By combining the current relative velocity with its change over the time step, the MLV improves friction-branch discrimination near zero relative velocity in this case.

### *4.2. Effect of the friction-velocity slope on computational performance*

In velocity-dependent friction models, a decrease in the friction coefficient with increasing relative sliding speed gives rise to a negative friction-velocity slope. A steeper negative slope near zero sliding speed increases the sensitivity of the friction force to velocity perturbations and may promote stick-slip instability [47, 48]. To examine its effect on prediction accuracy, two friction conditions are considered for Case III. The baseline condition uses $\alpha = 2$, whereas $\alpha = 1000$ produces a sharper decrease in the friction coefficient near zero sliding speed, as shown in Fig. 15(a).

Fig. 15(b) compares the RMSE values for $\alpha = 2$ and $\alpha = 1000$. Each condition is evaluated against its own independently computed benchmark result. The RMSE values for $\alpha = 2$ and $\alpha = 1000$ remain generally of a similar order of magnitude, indicating that the steeper negative slope near zero sliding speed does not substantially increase the response errors in this test.

Fig. 15(c) shows the total loss at successive time steps. For $\alpha = 1000$, pronounced local peaks reach the order of $10^{-4}$ or higher, whereas for $\alpha = 2$, the loss is predominantly of the order of $10^{-12}$. The steeper negative slope therefore produces larger local loss values at some time steps. Together with Fig. 15(b), these results indicate that the local loss peaks do not lead to a substantial increase in the overall response errors in this test.

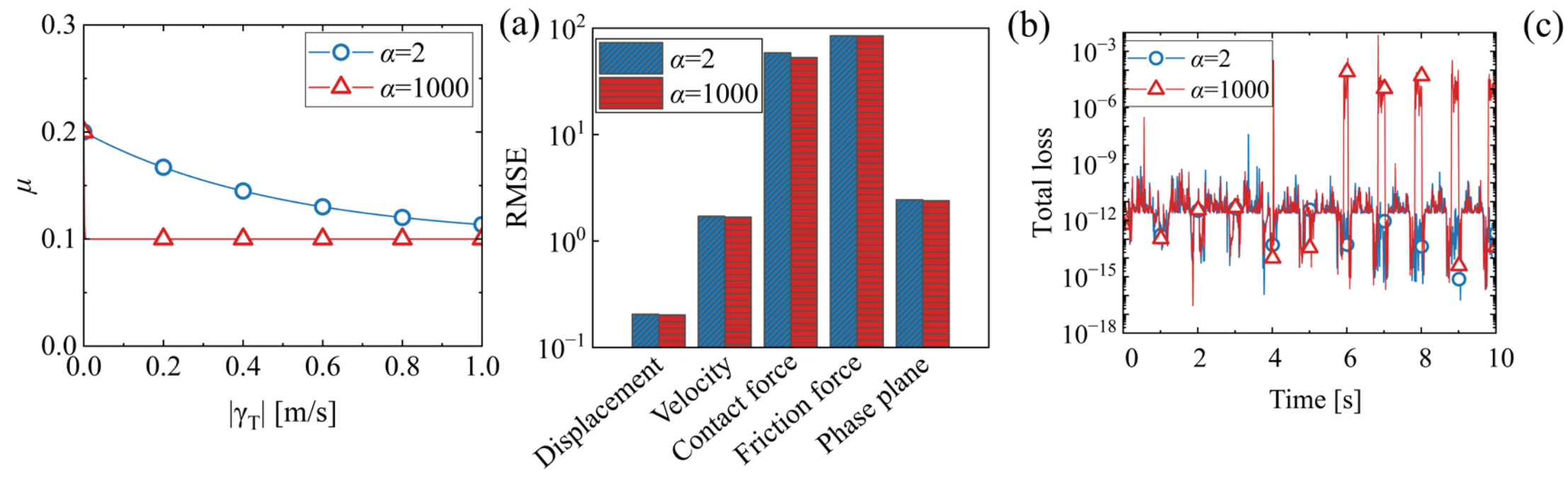

Fig. 15. Effect of the negative friction-velocity slope on computational performance: (a) friction-coefficient curves; (b) RMSE comparison; (c) total loss over time.

*4.3. Long-time error growth and time-frequency characteristics*

In nonsmooth systems, errors in switching times can cause phase drift and changes in time-frequency characteristics, in addition to local amplitude errors. To examine long-time error growth, the simulation of Case III is extended from 10 s to 60 s. The RK4-LCP method defined in Section 3.4.2, with a time step of $\Delta t = 10^{-5}$, provides the benchmark result. The errors of RK4-LCP with $\Delta t = 10^{-4}$ and MLV-PINN with $\Delta t = 10^{-2}$ are evaluated against this benchmark result.

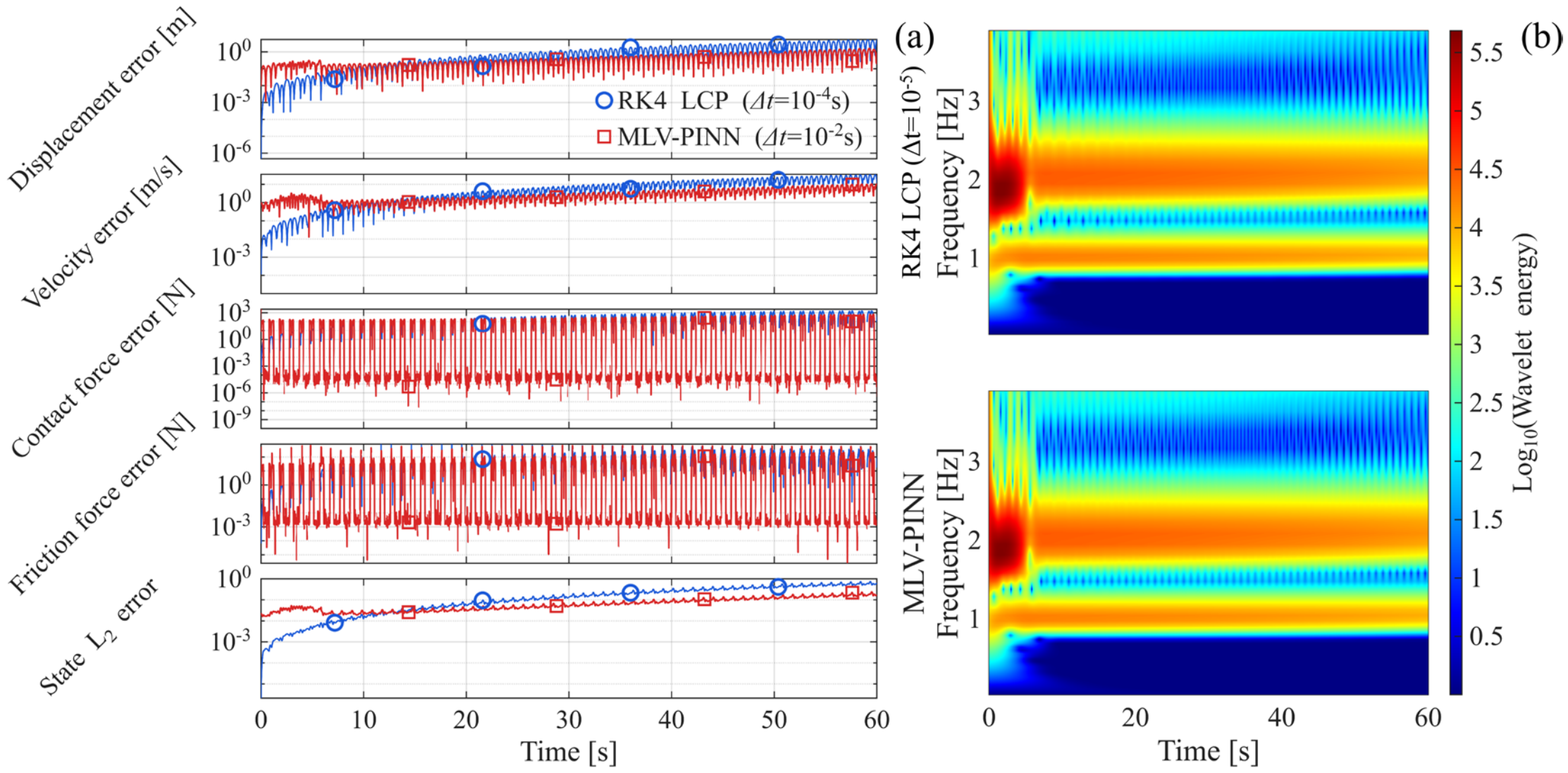

Fig. 16. Long-time error evolution and time-frequency characteristics: (a) mean displacement error, mean velocity error, normal contact-force error, friction-force error, and normalised instantaneous state-vector $L_2$ error; (b) wavelet time-frequency spectra of the horizontal acceleration obtained from the RK4-LCP benchmark result $(\Delta t = 10^{-5})$ and MLV-PINN.

Fig. 16(a) presents the mean displacement and velocity errors, normal contact-force error, friction-force error, and instantaneous state-vector $L_2$ error over 60 s. The mean displacement and velocity errors are averaged over the horizontal and normal degrees of freedom. For the instantaneous state-vector $L_2$ error, the displacement and velocity deviations are first normalised by the maximum absolute values of the corresponding components of the high-fidelity benchmark result, and the $L_2$ norm of the resulting dimensionless displacement-velocity error vector is then evaluated. During the first 10 s, RK4-LCP yields lower errors than MLV-PINN for most metrics, consistent with its smaller time step. Over approximately 20-60 s, its errors increase and exceed those of MLV-PINN for several metrics, particularly displacement, velocity, and the normalised instantaneous state-vector $L_2$ error. MLV-PINN has slightly larger initial errors but slower subsequent growth, leading to lower errors in these quantities for the long-time simulation than RK4-LCP with $\Delta t = 10^{-4}$.

Fig. 16(b) compares the wavelet time-frequency spectra of the horizontal acceleration obtained from the RK4-LCP benchmark result $(\Delta t = 10^{-5})$ and MLV-PINN, illustrating the transient response associated with contact-friction transitions. The dominant energy bands and their evolution over time agree closely, indicating that MLV-PINN preserves the principal frequency content in this long-time test.

*4.4. Sensitivity to loss weights and Adam pre-training steps*

The residual terms in the composite physics-informed loss function in Eq. (9) differ in scale and gradient characteristics, and inappropriate loss weights may bias optimisation towards particular constraints [29]. Case III, which includes both normal contact-separation and tangential stick-slip transitions, is used to examine sensitivity to the loss weights and the number of Adam pre-training steps. All residuals are nondimensionalised using their characteristic scales before the losses are calculated, so the weights specify the relative contributions of the resulting dimensionless loss terms. Nine configurations are tested using Adam pre-training followed by L-BFGS-B refinement. Loss weights are specified separately for the Adam and L-BFGS-B stages, and the number of Adam pre-training steps is varied. The other L-BFGS-B optimisation parameters are kept fixed. All RMSE values are evaluated against the same benchmark result.

Table 2 lists the nine training configurations. G0 is the baseline used in the preceding examples; G1-G6 vary the loss weights, while G7 and G8 use fewer and more Adam pre-training steps, respectively. Table 3 reports the RMSE values, and Fig. 17 shows their percentage changes relative to G0.

Table 2. Nine training configurations used in the sensitivity test for Case III.

| Group | Adam stage weights ($w_{\text{dyn}}, w_{\text{N}}, w_{\text{T}}, w_{\text{cons}}$) | L-BFGS-B stage weights ($w_{\text{dyn}}, w_{\text{N}}, w_{\text{T}}, w_{\text{cons}}$) | Adam steps |
|---|---|---|---|
| G0 | (1, 10, 20, 1) | (2, 20, 200, 0.5) | 4000 |
| G1 | (1, 1, 1, 1) | (1, 1, 1, 1) | 4000 |
| G2 | (1, 0.1, 0.1, 1) | (1, 0.1, 0.1, 1) | 4000 |
| G3 | (1, 1, 1, 0.1) | (1, 1, 1, 0.1) | 4000 |
| G4 | (1, 1000, 1, 1) | (1, 1000, 1, 1) | 4000 |
| G5 | (1, 1, 1000, 1) | (1, 1, 1000, 1) | 4000 |
| G6 | (1000, 1, 1, 1) | (1000, 1, 1, 1) | 4000 |
| G7 | (1, 10, 20, 1) | (2, 20, 200, 0.5) | 1000 |
| G8 | (1, 10, 20, 1) | (2, 20, 200, 0.5) | 7000 |

Table 3. RMSE values obtained by each training configuration for Case III with respect to the high-fidelity numerical benchmark result.

| Group | Disp. RMSE | Vel. RMSE | Fric. RMSE | Cont. RMSE | Phase RMSE |
|---|---|---|---|---|---|
| G0 | 0.20476 | 1.7077 | 58.498 | 84.677 | 1.7204 |
| G1 | 0.20480 | 1.7079 | 58.496 | 84.687 | 1.7206 |
| G2 | 0.20480 | 1.7079 | 58.498 | 84.685 | 1.7206 |
| G3 | 0.20479 | 1.7079 | 58.496 | 84.685 | 1.7206 |
| G4 | 0.20478 | 1.7078 | 58.498 | 84.681 | 1.7205 |
| G5 | 0.20486 | 1.7085 | 58.505 | 84.709 | 1.7212 |
| G6 | 0.20684 | 1.7243 | 60.276 | 85.331 | 1.7371 |
| G7 | 0.20478 | 1.7078 | 58.500 | 84.680 | 1.7205 |
| G8 | 0.20476 | 1.7077 | 58.497 | 84.677 | 1.7204 |

Table 3 and Fig. 17 show that, except for G6, the RMSE values differ from those of G0 by no more than approximately 0.05%. In G6, increasing the equations-of-motion residual-loss weight $w_{\mathrm{dyn}}$ to $1000$ raises all RMSE values, with the largest increase of 3.039% occurring in the friction-force RMSE. This suggests that excessive emphasis on the equations-of-motion residual $\mathcal{L}_{\mathrm{dyn}}$ can disturb the balance among the physical constraints and reduce the overall prediction accuracy. Reducing the Adam pre-training steps to 1000 in G7 or increasing them to 7000 in G8 produces RMSE values close to those of G0.

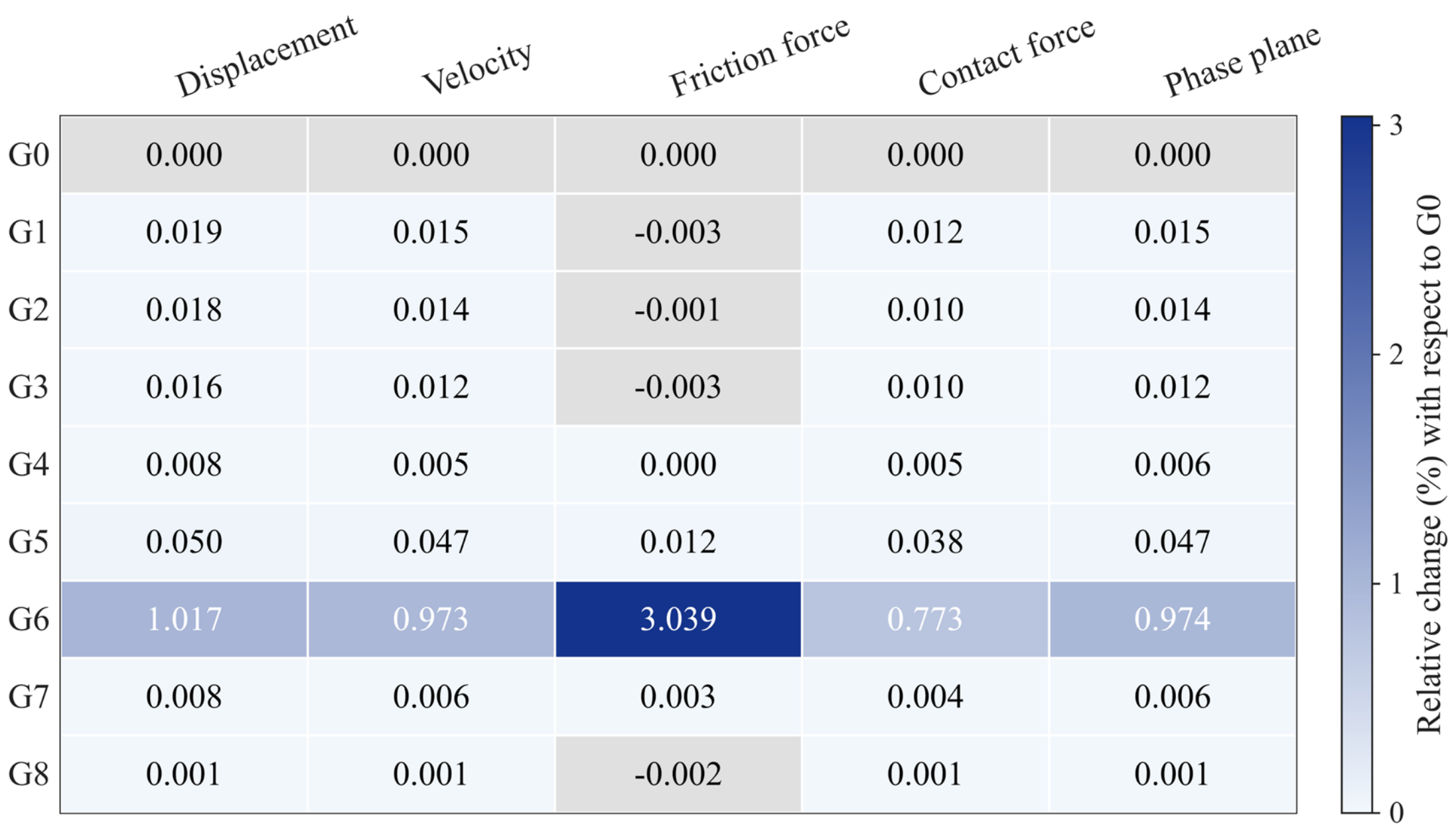

Fig. 17. Heatmap of the percentage change in RMSE for each group relative to the baseline configuration G0.

These results indicate that several training configurations achieve accuracy comparable to G0 within the tested range, provided that the equations-of-motion residual loss is not excessively weighted.

## 5. Conclusions

This study proposes a general PINN framework for high-accuracy computation of nonsmooth dynamics in complex systems with multiple frictional contacts, which is termed the mixed leading variable physics-informed neural network (MLV-PINN). Inspired by the physical interaction between system motion and frictional contact, it represents motion and contact variables as independent network outputs and determines them jointly under the governing equations and contact-friction constraints. It provides a common solution framework for both discrete systems governed by ordinary differential equations and continuum systems originally governed by partial differential equations. The main conclusions are as follows:

(1) In terms of physical loss function design, a composite loss combines the original second-order equations of motion, contact-friction complementary conditions, and friction-margin consistency relation, enabling dynamic responses and contact states to be determined jointly under physical constraints. The smoothed Fischer-Burmeister formulation incorporates the complementary conditions into network optimisation without separately imposing non-negativity and orthogonality penalties.

(2) To improve the contact-state identification criterion in PINN, this study introduces a mixed leading variable based on the relative tangential velocity and acceleration and uses it to reformulate the tangential complementary conditions, thereby improving friction-branch discrimination near zero relative velocity. The ablation study shows more accurate stick-slip transition times and associated friction-force changes than PINNs using relative velocity or acceleration alone.

(3) In terms of network architecture design, MLV-PINN treats accelerations, normal contact forces, and friction margins as independent network outputs. Together with the direct physical constraints, this design reduces the need for intermediate algebraic reformulation and variable elimination in constructing discrete velocity-impulse complementary subproblems. To achieve high computational accuracy, a high-order implicit Runge-Kutta scheme reconstructs velocities and displacements from the stage accelerations predicted by the network, while maintaining the discrete kinematic relations among the motion variables.

(4) The framework is tested on four nonsmooth problems covering tangential stick-slip, multiple coupled contacts, coupled normal contact-separation and tangential stick-slip, and moving contact on an elastic beam. The first three cases indicate the effectiveness of the framework for one-dimensional stick-slip motion and two-dimensional motion involving both tangential and normal contact-state transitions. A solution procedure is further developed for the continuum problem in Case IV. Modal discretisation allows the beam response to be represented by generalised coordinates, so the same MLV-PINN framework can be retained by incorporating the modal equations of motion and the contact relations at the moving contact location into the physical loss.

(5) MLV-PINN exhibits slower error growth during the later response and preserves the principal time-frequency features. At the relatively large time steps tested, MLV-PINN reduces displacement and velocity RMSEs by 85.16%-99.95% relative to the corresponding conventional methods. A steeper negative friction-velocity slope produces response errors of a similar order of magnitude, despite larger loss peaks at some time steps. Variations in the loss weights and Adam pre-training steps also yield errors close to the baseline, except when the equations-of-motion residual loss is excessively weighted.

The proposed framework offers a promising approach to the computation of nonsmooth dynamics. Future work will explore further ways to combine neural network representations with the underlying physics of contact and friction.

**CRediT authorship contribution statement**

**Xinyuan Li:** Methodology, Software, Validation, Investigation, Data curation, Writing - original draft, Visualization. **Zilin Li:** Conceptualization, Formal analysis, Resources, Writing - review & editing, Supervision, Funding acquisition. **Hongtao Wei:** Software, Supervision. **Ronghan Wei:** Resources, Writing - review & editing, Project administration.

**Data availability**

The datasets generated during the current study are not publicly available as the basic equations and information have been clearly presented, but are available from the corresponding author on reasonable request.

**Declaration of competing interest**

The authors declare that they have no known competing financial interests or personal relationships that could have appeared to influence the work reported in this paper.

**Acknowledgments**

The authors acknowledge financial support received from the National Natural Science Foundation of China (Grant No. 12402242).

**Declaration of generative AI use**

The entire manuscript was originally written by the authors. ChatGPT was used solely to check grammar and improve language and readability. After using this tool, the authors reviewed and edited the content as necessary and take full responsibility for the content of the published article.